\documentclass[12pt]{article}
\usepackage{amssymb, amsmath, latexsym, multicol, multirow, epsfig, color, subfigure, setspace, array, amsbsy, multirow, amssymb, color, amsbsy, epsfig, verbatim, diagbox}
\usepackage[stable]{footmisc}
\usepackage[authoryear]{natbib}
\usepackage[latin1]{inputenc}
\usepackage{tikz}
\usetikzlibrary{shapes,arrows}
\usetikzlibrary{positioning,calc}
\usepackage{float}

\title{Using particle swarm optimization to search for locally $D$-optimal designs for mixed factor experiments with binary response}
\date{\today}
\author{Joshua Lukemire \\ Emory University \and
Abhyuday Mandal \\ University of Georgia \and
Weng Kee Wong \\ University of California at Los Angeles}

\begin{document}
\renewcommand{\baselinestretch} {1.06}

\maketitle
\begin{abstract}
Identifying optimal designs for generalized linear models with a binary response can be a challenging task, especially when there are both continuous and discrete independent factors in the model. Theoretical results rarely exist for such models, and the handful that do exist come with restrictive assumptions.   This paper investigates the use of particle swarm optimization (PSO) to search for locally $D$-optimal designs for generalized linear models with discrete and continuous factors and a binary outcome and demonstrates that PSO can be an effective method. We provide two real applications using PSO to identify designs for experiments with mixed factors: one to redesign an odor removal study and the second to find an optimal design for an electrostatic discharge study.  In both cases we show that the $D$-efficiencies of the designs found by PSO are much better than the implemented designs.  In addition, we show  PSO can efficiently find $D$-optimal designs on a prototype or an irregularly shaped design space,  provide insights on the existence of  minimally supported optimal designs, and evaluate sensitivity of the $D$-optimal design to  mis-specifications in the link function.
\end{abstract}



\section{Introduction}

Our work is motivated by an odor removal study (Wang et al., 2015) conducted in the Department of Textiles, Merchandising, and Interiors within the College of Family and Consumer Sciences at the University of Georgia (UGA). In the research, algae were used in the creation of bio-plastics due to their high protein content. Algae are preferable to other products such as soy because they are easily acquired as an agricultural waste product and are not as in demand for other uses. In bio-plastic formulation, the algae must undergo some chemical processing to force their proteins to have the desired properties. As a side effect of this chemical processing, algae-based bio-products will often have unpleasant odors. These odors must be removed or at least diminished if the products are to be used for commercial purposes. Table 1 shows the discrete and continuous factors deemed important in the study, but the researchers at UGA implemented a design to estimate parameters in a model with a binary response assuming a constant temperature of 25$^\circ$C.  A main reason for artificially fixing the temperature level was that the researchers were not able to find an efficient design when they included temperature as a continuous factor in the model.

\begin{table}
\caption{\label{biofactors} Factor types and levels for the bio-plastics odor experiment. \vspace{0.2in}}
  \centering
  \makebox{
\begin{tabular}{|c|c|cc|}
\hline
Type      & Factor  & \multicolumn{2}{c|}{Levels} \\
&         &  $-$ & $+$ \\
\hline
\multirow{4}{*}{Discrete}         & Algae   & Catfish algae &  Solix Microalgae\\
 & Scavenger  &  Activated Carbon &  Zeolite\\
& Resin      & Polyethylene & Polypropylene \\
&  Compatibilizer   & Absent & Present \\
\hline
Continuous & Temperature & \multicolumn{2}{c|}{Temperature from 5$^\circ$C to 35$^\circ$C}\\
\hline
\end{tabular}
  }
\end{table}

We propose using particle swarm optimization as a tool for finding optimal designs for models with both discrete and continuous factors and binary outcome. Similar to the experiment conducted in Wang et al. (2015), the response can be whether or not the odor is successfully removed, but we now include all five factors. The first four factors are algae type, scavenger material, type of resin, and compatabilizer, each taking two levels. The fifth factor is the storage temperature of the algae, which is continuous and is now allowed to range from 5$^\circ$C to 35$^\circ$C (see Table \ref{biofactors}). The assumed model is  $logit(\mu) = \beta_0 + \beta_1 Algae + \beta_2 Scavenger + \beta_3 Resin + \beta_4 Compatibilizer + \beta_5 Temperature$ and the goal is to estimate all parameters in the model, where $\mu$ is the probability of an unpleasant odor in the bio-plastic.

Khuri et al. (2006) provides an overview and background on design issues for generalized linear models. Analytical derivation of the optimal designs for such models typically assumes that there is  a single factor or that the factors are additive when there are several of them.  A common setup is that the model has all discrete factors or all continuous factors.  For example, \cite{multvar} provides optimal designs for experiments when all factors are continuous and \cite{factorial2k} provides optimal designs for studies when all factors are discrete. When theoretical results are not available, computational methods are used to find optimal designs. A recent overview of algorithms for finding optimal designs is Mandal et al. (2015). Specific applications of computational methods to solve real design problems can be found in Woods et al. (2006), Dror and Steinberg (2006, 2008), Waterhouse et al. (2008), and Woods and van de Ven (2011).  The numerical techniques employed in these papers require a user-selected candidate set of design points and the search algorithm finds the optimal weight distribution for these points.  When the number of continuous factors increases, the size of this candidate set quickly grows and the computational effort can be prohibitively expensive.

There is little work on constructing efficient designs when there are both discrete and continuous factors.  Such studies with mixed factors pose special challenges both from the theoretical and computational standpoints.  A main reason is that the theory for constructing optimal designs when all factors are continuous or when all factors are discrete does not extend to the case when there are mixed factors.  Consequently, there are no efficient algorithms that we know of for finding optimal designs for such models.  There appears to be increasing interest in finding efficient designs for experiments with mixed factors and recent work in this direction for computer experiments includes Deng et al. (2013) and Huang et al. (2015).

Our primary goal is to identify locally $D$-optimal designs for generalized linear models with a binary response having  both continuous and discrete independent factors. The main aim of this paper is to show that particle swarm optimization (PSO) techniques can be useful for finding optimal designs for such models and therefore fill an important gap in the literature.  PSO is a nature-inspired metaheuristic algorithm that has been widely used in engineering and computer science to tackle general optimization problems successfully.  As we will show, PSO is  able to generate optimal designs for studies with both discrete and continuous factors efficiently. It does so without making assumptions on the problem and also does not require a candidate set of points to be specified in advance.  In what is to follow, we apply PSO to search for a variety of optimal designs for binary response GLMs with mixed factors. We show our numerical results coincide with the limited results in the literature obtained analytically using a restrictive assumption.  We then revisit our motivating problem and  show PSO finds an optimal design for the odor removal study with mixed factors that is nearly twice as efficient as the one implemented by the researchers.

 This paper is organized as follows. Section 2 describes background and briefly reviews previous design work for generalized linear models with a binary response and factors that are either all continuous or all discrete. Section 3 introduces PSO and describes how it can be used to search for optimal designs  in a general design problem. In Section 4, we apply PSO to find optimal designs for generalized linear models with binary response when all  factors are discrete or continuous and also when there are mixed discrete and continuous factors. We revisit our motivating odor removal example in Section 5 and show that PSO can produce a design much more efficient than the design used in practice.  We also work out another similar application from the literature to design an electrostatic discharge study.  Section 6 further demonstrates the flexibility of PSO for finding optimal designs for other models with a binary response, including cases when the design space is irregularly shaped. Section 7 summarizes our work with some concluding remarks.


\section{Optimal Designs for Generalized Linear Models}

A generalized linear model (GLM) is appropriate when the response $Y$ has a distribution from the Exponential family in canonical form.  Observations are independent and each response $Y_i$ has mean  $E(Y_i) = \mu_i$, related by a link function $g(.)$ to the linear predictor $\eta_i = x_i^T \beta$ by $g(\mu_i)=\eta_i$.  Here $x_i$ is the $i^{th}$ combination of factor levels used to observe $Y_i$ and the range of values for each factor is assumed known. The link function is monotonic and its form depends on the type of outcome. For example, if the outcome is binary, there are four typical link functions:

\[ g(\mu) = \left\{
  \begin{array}{l l}
    \log(\frac{\mu}{1-\mu}) & \quad \text{logit link}\\
    \Phi^{-1}(\mu) & \quad \text{probit link}\\
    \log(-\log(1-\mu)) & \quad \text{complementary log-log link}\\
    \log(-\log(\mu)) & \quad \text{log-log link}.
  \end{array} \right.\]

The goal of an experiment is usually to make inference on a treatment effect  corresponding to a selected parameter in the model or some meaningful function of the parameters in the model. The design problem is then to find the optimal number of combinations of factor levels and the types of combinations of factor levels that  minimizes the  variance of the estimated parameter or the generalized variance of the  estimated parameters of interest if there are several of them.  When the design $\xi$ has $m$ distinct combinations of the levels of the factors and there are $n_i$ replicates at each $x_i$, a direct calculation shows that the information matrix of the design $\xi$ is
\[
I_{\xi} = \sum_{i=1}^m n_i  \Psi{(\eta_i)}  x_i x_i^T,
\]
where  $\eta_i = x_i^T \beta$   and $\Psi(\eta_i) = \frac{(d\mu_i/d\eta_i)^2}{\mu_i(1-\mu_i)}$. For instance, when the link is the logit function, we have
\[
\Psi(\eta_i) = \frac{1}{2 + e^{\eta_i} + e^{-\eta_i} } = \frac{e^{\eta_i}}{(1+e^{\eta_i})^2}.
\]

A $D$-optimal design maximizes the log-determinant of the information matrix and so it is appropriate for estimating all parameters in the model because it minimizes the generalized variance. Because the matrix depends on the model parameters, nominal values of the parameters are required before we can implement the design. Such nominal values typically come from literature or prior experiences based on pilot studies. The resulting optimal designs are therefore locally optimal, which are often used as building blocks for constructing more complicated designs (Ford et al. 1992).

For a predetermined sample size $N$ and a design criterion, the design problem is to determine the optimal values of $m$, $x_1,\ldots, x_m$ and $n_1,\ldots,n_m$ subject to $n_1+\ldots+n_m=N$.  These are exact design problems and they are generally very difficult to solve. We resort to finding approximate optimal designs by not optimizing each $n_i$ directly but by optimizing $p_i = n_i/N$, the proportion of the total observations to be taken at $x_i,i=1,\ldots,m$ subject to the constraint that they sum to unity (Silvey, 1980). The resulting designs are called optimal approximate designs and are  implemented   by taking roughly $Np_i$ observations at $x_i$ after the $Np_i$'s are rounded to positive integers such that they sum to $N$.

When the design criterion is formulated as a convex function of the information matrix, we can verify the optimality of an approximate design  among all designs using an equivalence theorem, see for example, Kiefer and Wolfowitz (1959) or Pukelsheim (1993). For the logistic model with a logit link function and the $k\times 1$ vector of unknown parameters $\beta$ in the linear predictor, this theorem asserts that the design $\xi^*$ is $D$-optimal among all designs if and only if for all $x$  in the design space,
\begin{equation}
	\frac{e^{\beta^T x}}{(1+e^{\beta^T x})^2} \;  x^T \; I_{\xi^*}^{-1} \;  x - k  \leq 0,
\label{equiveqn}
\end{equation}
with equality at each support point of $\xi^*$.  The function to the left of the above inequality is sometimes called the sensitivity function.  When there is only one factor in the model, optimality of the design $\xi$ can be easily confirmed by plotting its sensitivity function across the design space and observing whether the graph has the desired properties.

Often the worth of a design is measured by its efficiency relative to the optimal design. For example, the $D$-efficiency of a design $\xi$ for a model with $k$ parameters in the linear predictor is
\begin{equation}
	\bigg{(}\frac{det (I_{\xi})}{det(I_{\xi^*})}\bigg{)}^{1/k}.
	\label{lowerbound}
\end{equation}

\noindent If the ratio is one half, then the design $\xi$ has 50\% $D$-efficiency, or in other words, this design will require twice as many replicates as the optimal design to get the same information. When the true optimum design is unknown, a lower bound on the $D$-efficiency of a design $\xi$  is $e^{-\theta / k}$, where $\theta$ is the maximum positive value of the sensitivity function across the design space (Pazman, 1986). Clearly, when $\xi$ is $D$-optimal $\theta = 0$, and the lower bound attains unity. In the next section, we discuss PSO as a general optimization tool and use a PSO-based algorithm to search for optimal designs. Our proposed algorithm incorporates the efficiency lower bound of the generated design as our stopping rule to terminate the search.

\section{Particle Swarm Optimization}
PSO is a metaheuristic optimization algorithm based on animal behavior introduced by Kennedy and Eberhart in 1995 (Kennedy and Eberhart, 1995). PSO has been used to solve a wide variety of optimization problems in several disciplines. For example, in public health research, Fu et al. (2009) used PSO to identify optimal screening nodes for spread of the SARS disease in Singapore; other applications are voluminously documented in the engineering literature, such as in the IEEE transactions. The use of PSO for finding optimal designs is relatively new but seems to be growing in popularity. For example, Qiu et al. (2014) used PSO to solve a variety of biomedical design problems including estimating parameters in compartmental models and tumor growth models. It has also been used to find optimal designs under a non-differentiable optimality criterion (Chen et al., 2014) and optimal designs for a variety of mixture models (Wong et al., 2015).  The major appealing features of PSO are that  the search space needs not be discretized, it is fast and easy to implement, and makes no assumption on the optimization problem it is trying to solve.  Our work here is the first to use PSO to find a variety of locally $D$-optimal designs for GLMs with mixed factors and also show PSO can find the optimal design when assumptions required for its theoretical construction are not valid.

PSO is a nature-inspired algorithm that mimics the behavior of a flock of birds as they search over an area for food. Each member of the flock or swarm, known as a particle, represents a candidate design solution to the problem we are trying to solve, and the location of the food represents the optimum solution. Each bird or particle has a fitness value; in our case, this value is  the determinant of the information matrix of the design (or bird). Each particle has its own perception of where the food is located based on its own past experience and this position is known as the personal best (PBest). Members in the swarm exchange information cognitively with one another while searching over the area and decide collectively where the food is located.  This position is called the global best position (GBest).  Consequently,  at each time point, the flock flies toward the GBest position and each particle is also being pulled in the direction of its PBest position at the same time.  These positions are updated at each iteration as the algorithm proceeds.

\subsection{Algorithm Overview}

\phantom{space}
\tikzstyle{eblock} = [rectangle, draw, fill=green!10, text width=2cm, text centered, rounded corners, minimum height = 2cm]
\tikzstyle{init} = [rectangle, draw, fill=white!10, text width=2cm, text centered, rounded corners, minimum height = 2cm]
\tikzstyle{fisher} = [rectangle, draw, fill=white!10, text width=3.5cm, text centered, rounded corners, minimum height = 2cm, minimum width = 3.7cm]
\tikzstyle{line} = [draw, -latex']
\tikzstyle{mblock} = [rectangle, draw, fill=green!10, text width=2cm, text centered, rounded corners, minimum height = 2cm]
\tikzstyle{loopstep} = [draw, rectangle, fill = white!10, node distance=3cm, minimum height = 2cm]
\tikzstyle{choice} = [draw, diamond, fill = green!10, node distance=3cm, minimum height = 2cm]

\begin{center}
\begin{tikzpicture}[node distance = 3.5cm, auto]

\node [init] (inits) {Initialize swarm};
\node [init, right of = inits] (upvel) {Update Velocity};
\node [init, right of = upvel] (uppos) {Update Position};
\node [fisher, right of = uppos] (calcfit) {Calculate determinant of the Fisher Information};
\node [init, below of = calcfit] (checkim){Check Improvement};
\node [init, left of = checkim] (converge) {Equivalence check required?};
\node [init, left of = converge] (equiv) {Pass Equivalence Theorem?};
\node [init, below of = equiv] (output) {Output Final Design};

\path [line] (inits) -- (upvel);
\path [line, dashed] (upvel) -- (uppos);
\path [line, dashed] (uppos) -- (calcfit);
\path [line, dashed] (calcfit) -- (checkim);
\path [line, dashed] (checkim) -- (converge);
\path [line, dashed] (converge) -- node{no}(upvel);
\path [line, dashed] (converge) -- node{yes}(equiv);
\path [line, dashed] (equiv) -- node{no}(upvel);
\path [line] (equiv) -- node{yes}(output);

\end{tikzpicture}
\end{center}

\begin{description}
\item [Initialize Swarm:] Generates the particle swarm. Each particle is given a random position in the search space, which corresponds to a random experimental design with $2^k$ design points. The $D$-optimality of each design is then calculated and stored as the Personal Best (PBest). Finally the maximum PBest among all particles is stored as the Global Best (GBest).

\item [Update Velocity:] This step calculates the distance each particle is from its personal best known position and the global best position. These two distances are combined with the previous velocity according to Equation~\ref{veleqn} (described later) to generate a new velocity.

\item [Update Position:] In the update position step each particle's position has its velocity added to it. If the velocity is greater than zero, this corresponds to a new experimental design. If this step results in a design that is out of bounds (for example a factor setting moves beyond its upper bound) then the particle is brought back to the boundary and its velocity of modified to reflect this change.

\item [Calculate Fitness:] Calculate the determinant of the Fisher Information for each candidate design (particle).

\item [Check Improvement:] Checks to see which particles have current fitness values that are better than their personal best value. This corresponds to a design that is better than any design that particle had discovered so far. Any particles where this is true have their PBest updated. Finally these PBests are checked to see if any are better than the current global best.

\item [Check Equivalence Theorem:] Check the global best design to see if it is $D$-optimal. This is done using Equation~\ref{equiveqn}. First, the design points included in the design are checked. If this check passes then the second check is run. For the second check a grid of values ranging from the lower to upper bounds of the continuous factors is used to generate alternative design points. These design points are also checked against Equation~\ref{equiveqn}. The maximum value obtained from running these points is stored as $\delta$ and the theoretical lower bound on the $D$-efficiency is calculated using Equation~\ref{lowerbound}. If this lower bound is high enough then the search terminates, otherwise the search continues. In our work, our stopping criterion was to terminate the search if the lower bound was greater than $99\%$.

\end{description}

\subsection{Tuning}
The particle swarm algorithm revolves around the way each particle's velocity is calculated. At each iteration every particle's velocity is updated based on its previous velocity, its personal best position, and the global best known position. The strength of each of these effects can be controlled by adjusting the inertia factor, cognitive learning factor, and the social learning factor. The inertia factor determines how much of the particle's velocity carries over from iteration to iteration. The cognitive learning factor and social learning factor modulate the effect of the distance from PBest and GBest respectively. The velocity is updated during each iteration according to Equation~\ref{veleqn}

\begin{equation}
\label{veleqn}
 v_{t} = \delta v_{t-1} + \varphi_1 U(0,1) (PBest - Pos) +  \varphi_2 U(0,1) (GBest - Pos),
\end{equation}

\noindent where $\delta$ is the inertia factor relating current velocity to previous velocity, $\varphi_1$ is the cognitive learning factor, and $\varphi_2$ is the social learning factor. When first introducing PSO, Kennedy and Eberhart (1995) used $\varphi_1 = \varphi_2 = 2$. We found these values to perform well for finding optimal designs. If one desires to increase the effect of one over the other, then the corresponding term can be increased. For example, if a higher emphasis on individual particle knowledge is desired, the cognitive learning factor might be increased, encouraging the particles to spread out more based on their own experience.

The magnitude of the inertia factor will have a powerful effect on the behavior of the swarm, especially the convergence of the particles. For our implementation of the algorithm we use a decreasing inertia with $\delta$ starting at 0.9 and decreasing by 0.01 every iteration until it reaches 0.4. This allows the particles to move very quickly soon after initialization, but forces them to slow down somewhat after they have had an opportunity to explore the search space.

Other than the tuning parameters related to the velocity updating step, the primary aspect of the algorithm to be tuned is the flock size, or the number of particles in the swarm. A smaller flock size will generally converge to a solution more quickly than a larger flock, but this comes at the risk of not exploring the search space as well. For identifying $D$-optimal designs, flocks as small as $5$ to $10$ particles appear to be adequate when the number of parameters is small (such as two or three), but as the number of parameters increases it becomes beneficial to increase the flock size.

The convergence tolerance and the minimum efficiency bound must be chosen based on the problem at hand. The convergence tolerance is how close the worst position must be to the global best position before the swarm is said to converge. Smaller values of the convergence tolerance will require the swarm to be closer but may increase runtime if the swarm has converged to a poor design. The minimum efficiency bound is calculated according to Equation~\ref{veleqn} and the minimum bound of the relative efficiency of the identified design to the theoretical $D$-optimal design. Generally we use a value of 99\% but this can be increased if the experimenter wishes for a tighter bound.


\section{PSO-generated Optimal Designs }

In this section we evaluate PSO's ability to generate locally $D$-optimal designs for a GLM with mixed factors and binary response. We do so by first briefly  showing  PSO can find optimal designs that coincide with those reported in the literature when all factors are discrete or all are continuous.  We also compare PSO with current algorithms and show that it has clear advantages over them, especially when all factors are continuous. In the last subsection, we apply PSO to generate locally $D$-optimal designs with mixed factors for other link functions and assess robustness properties of PSO-generated designs under a mis-specified link function. Throughout, we use the term resolution  to refer to how fine the equally spaced grid size is in the search space for algorithms that require it to be discretized. Because PSO does not require a discretized search space, we do not specify the resolution used to find the optimal design. Our experience is that results from PSO are usually consistent to more than two decimal places when it reports the support points and the weight distribution of the generated design. We also use the term reset to mean the maximum number of times we allow PSO to restart after failing to find a design that meets the lower efficiency bound before we terminate the algorithm and return the best result. For example, with a 200 reset number, the algorithm will restart up to 200 times if the maximum number of iterations is reached or the swarm converges to a bad design. If PSO has still not found a design that meets the user-specified efficiency lower-bound as a stopping criterion after 200 resets, it will stop resetting and will instead return the best design found during all searches.

All computations in this paper were carried out using a 2012 Macbook Pro 2.6GHz Intel Core i7 with 16G RAM on 64bit OSX Mavericks. Our code is written in C++ and called from R via the RCPP package (Eddelbuettel and Fran\c{c}ois, 2011). The C++ and R code for finding optimal designs in this paper can be downloaded from \\ {http://faculty.franklin.uga.edu/amandal/content/research}.

\subsection{All Discrete Factors}

First, we evaluate PSO's performance when all factors are qualitative and compare its designs to designs obtained using a few well-established algorithms for constructing $D$-optimal designs when all factors are discrete. A natural algorithm to compare with is the one most recently proposed by Yang et al. (2016), where they presented optimal designs for $2^k$ factorial experiments using a powerful and fast algorithm called Lift-One for finding designs when all factors are qualitative. We use the Lift-One algorithm and common ones like the Fedorov-Wynn algorithm, Cocktail algorithm, and the Multiplicative algorithms to generate designs and compare them with the designs found by PSO.

The model of interest is $logit(\mu) = \beta_0 + \sum_{i=1}^k \beta_i x_i$  for $k = 2,3,4$ with $x_i \in \{-1, 1\}$ and $\beta_0, \beta_i \sim U[-3,3]$. For each of the $2^2$, $2^3$, and $2^4$ experiments, we generate 10,000 optimal designs for a total of 30,000 designs per algorithm. All designs obtained are locally optimal designs and each is found using nominal values generated from the uniform distribution. Our goal is to compare the time it takes the algorithms to find all 10,000 designs for each of the $2^2$, $2^3$, and $2^4$ experiments as well as to examine the number of support points in the optimal designs.

We applied PSO to search for the $D$-optimal designs using 3, 8, and 25 particles for the $2^2$, $2^3$, and $2^4$ experiments respectively. PSO was limited to 100 maximum iterations and 200 maximum resets (although in practice far fewer resets are generally required). The swarm convergence tolerance was 0.00001 and the lower efficiency bound required to terminate PSO was 99\%. The Lift-One, Fedorov-Wynn, Cocktail, and Multiplicative algorithms were all run with the same convergence tolerance. Lift One was run with 100 maximum iterations. The Fedorov-Wynn Algorithm was run with 1000 maximum iterations and nonrandom initial points. The Cocktail and Multiplicative Algorithms were run with 1000 maximum iterations. The R implementation of the Fedorov-Wynn algorithm is based on \cite{fedorov2015}. The implementations of the Cocktail and Multiplicative algorithms follow \cite{yu2010} and \cite{algoopt}. The Lift-One algorithm is developed and implemented in \cite{factorial2k}.

Table~\ref{speed2k} displays the CPU times (in seconds) required by the algorithms to find all 10,000 optimal designs for each of the $2^2$, $2^3$, and $2^4$ experiments. Our results suggest that PSO is competitive with the others in terms of speed. The Lift-One algorithm is able to achieve faster times, which may not be surprising because the algorithm was specifically design for solving such problems. PSO is a general purpose algorithm and like all other algorithms it is not best for all situations, including situations when all factors are qualitative.

\begin{table}
 \caption{\label{speed2k} CPU time (seconds) to identify 10,000 main-effects only discrete factor optimal designs with nominal values generated from U(-3,3). \vspace{0.2in}}
  \centering
  \makebox{
  \begin{tabular}{|c|ccccc|}
  \hline
  & Lift-One & Fedorov-Wynn & Multiplicative & Cocktail & PSO \\
  \hline
  $2^2$ & 24.918 & 639.016 & 30.849 & 43.275 & 10.347 \\
  $2^3$ & 20.446 & 778.554 & 83.243 & 56.547 & 39.391 \\
  $2^4$ & 62.743 & 1558.144 & 598.634 & 440.214 & 239.350 \\
  \hline
  \end{tabular}
  }
\end{table}

\begin{figure}[h]
\begin{center}
\includegraphics[width=1.5in]{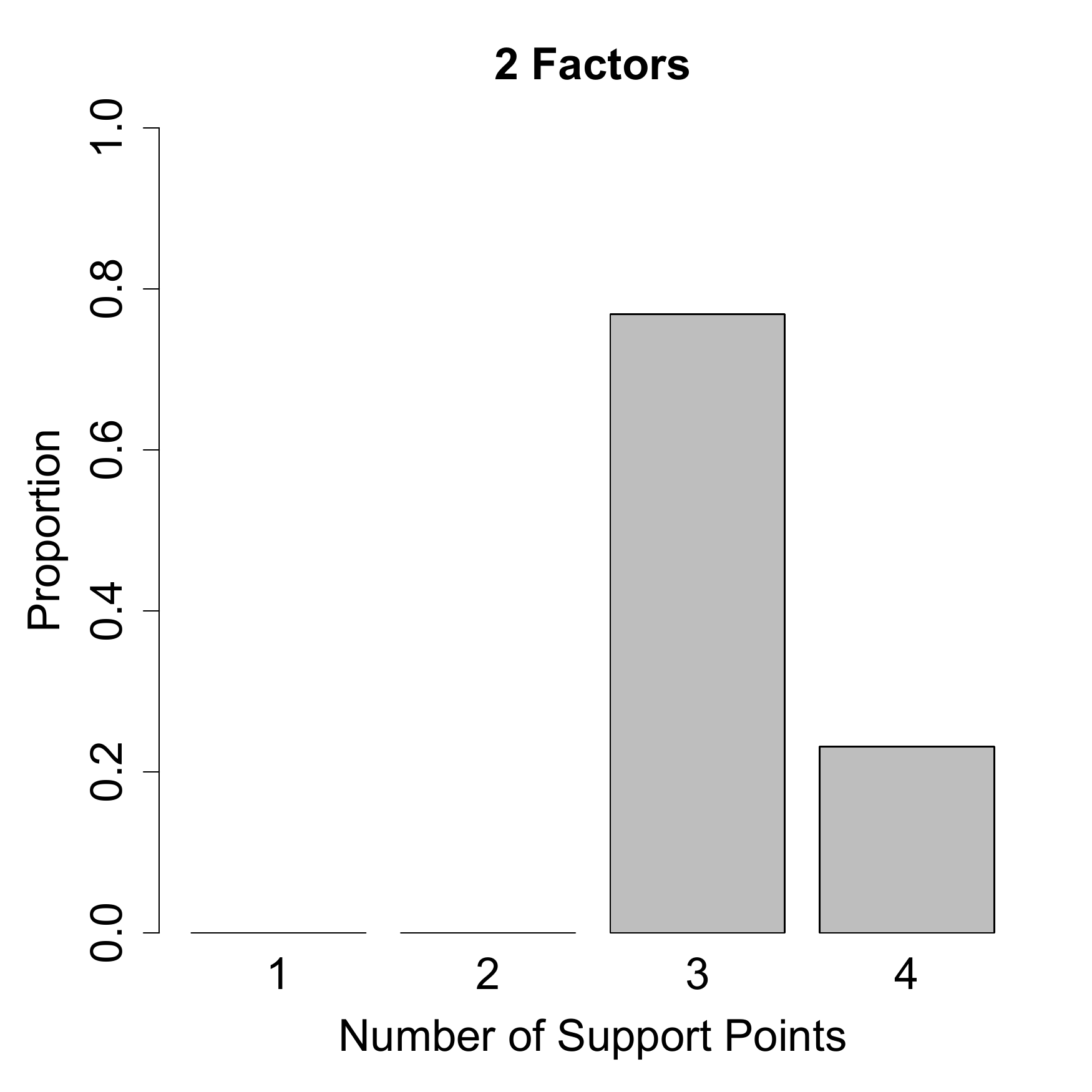}
\includegraphics[width=1.5in]{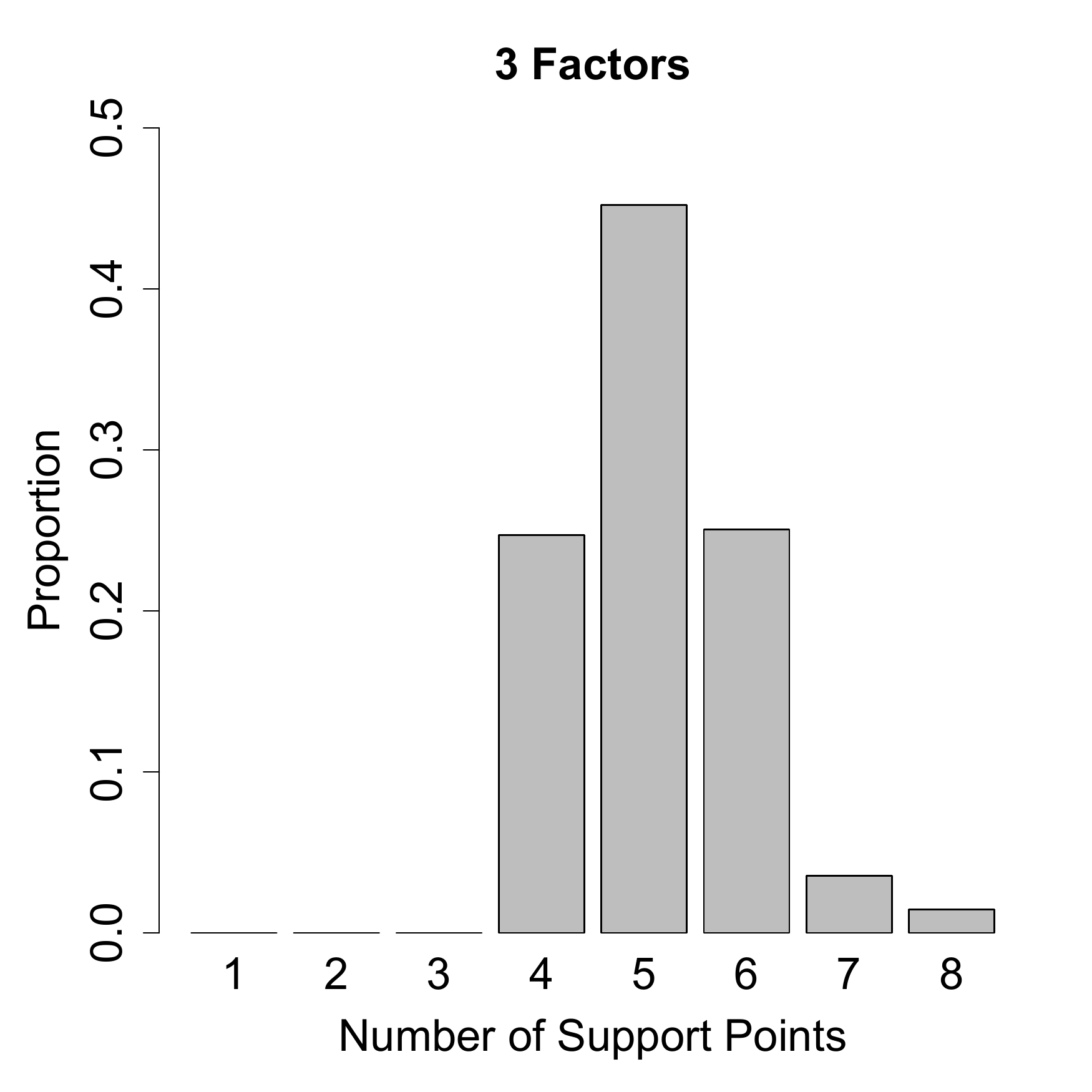}
\includegraphics[width=1.5in]{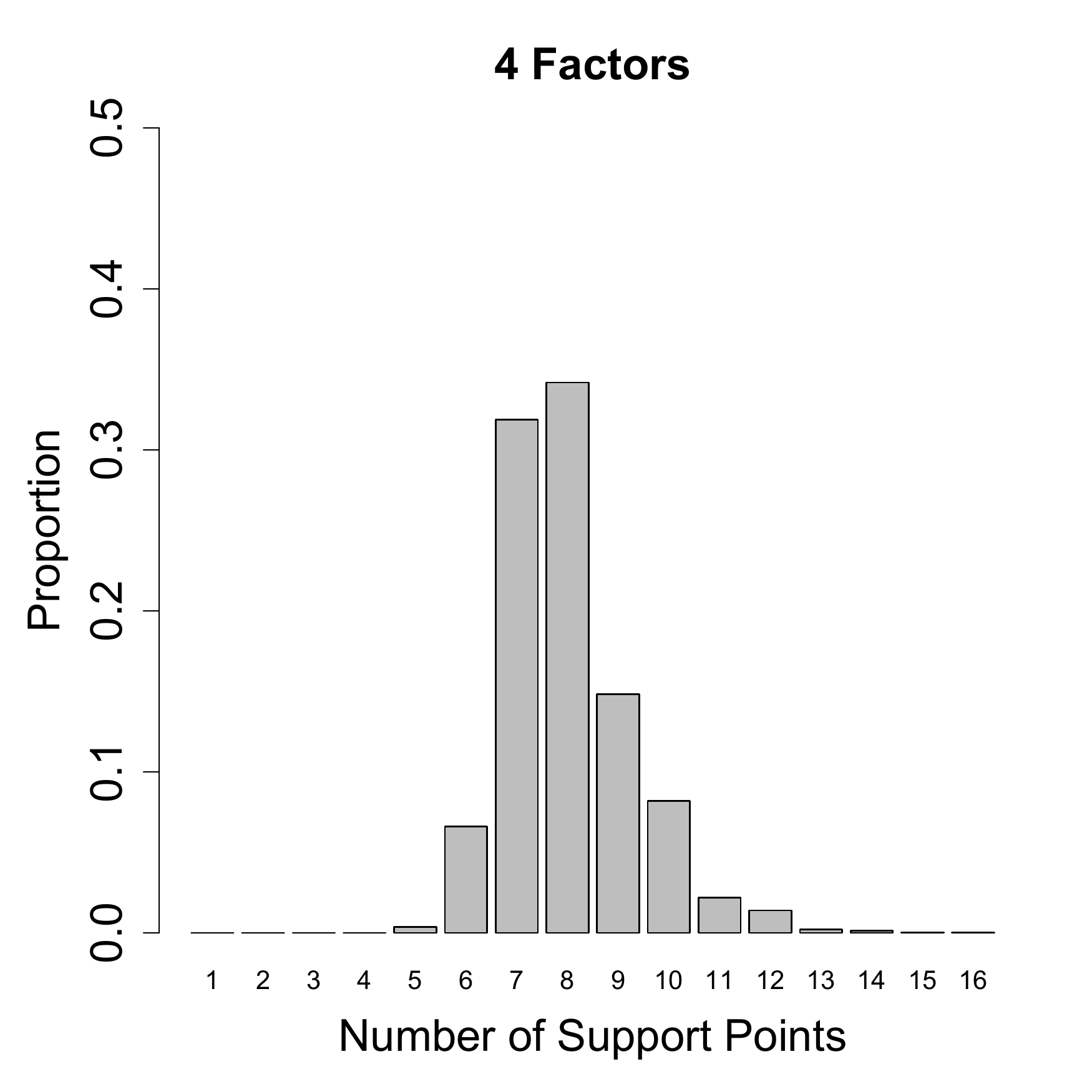}
\caption{Number of support points as the number of factors increases.}
\label{factorialsuppoints}
\end{center}
\end{figure}

We note that PSO is able to set the proportion allocated to a design point to zero, resulting in a smaller number of support points than the designs found by the Fedorov-Wynn and Multiplicative algorithms. It finds the same number of design points as the Lift-One algorithm, indicating that it is not finding designs that are any better than the designs found by the Lift-One algorithm (compare with Figure 3.1 in Yang et al. (2016)). In Figure~\ref{factorialsuppoints} we observe that as the number of factors increases the number of support points of the optimal design is no longer close to $2^k$. Frequently changing experimental settings can be costly, so an algorithms ability to identify designs with smaller numbers of support points can be advantageous from a cost perspective. Qiu et al. (2014) has also shown that another advantage of PSO over current algorithms is that it can also find singular optimal designs with ease by automatically dropping extra points during the search.


\subsection{All Continuous Factors}

PSO is particularly suited to explore a search space without requiring the space be discretized.  This is quite different from other algorithms discussed here, which require the user to specify a grid over the search space first. This can be problematic because the number of candidate points quickly becomes prohibitively large as the   range of values for each factor increases. We demonstrate the usefulness of PSO in comparison to these other algorithms by obtaining locally optimal designs for experiments with two continuous factors. The model of interest is $logit(\mu) = \beta_0 + \beta_1 x_1 + \beta_2 x_2$ with $x_1, x_2 \in [-1,1]$ and $\beta_0, \beta_1, \beta_2 \sim U(-3,3)$. Our procedure is to first apply PSO to obtain 1000 designs. The parameters for PSO were 15 particles, 100 maximum iterations, 100 maximum resets, convergence tolerance 0.00001, and a minimum lower efficiency bound of 99\%. We then obtain 1000 designs from each of the Lift-One, Fedorov-Wynn and Cocktail algorithms at a resolution of 0.1, 0.05, and 0.01.

Table \ref{cont_speed_sims} displays our simulation results. We observe that the CPU time required by the Cocktail algorithm is competitive with PSO when the resolution is 0.1, but  as the  resolution increases, the Cocktail algorithm requires significantly longer times to find the optimal design. Additionally, the previously mentioned lack of an ability to set the proportion allocated to a support point to 0 for the Cocktail and Fedorov-Wynn algorithms results in a likely impossible number of support points for their designs (e.g. with a resolution of 0.1 the Cocktail algorithm generated designs with 441 support points). Between its increased speed and ability to find designs with a smaller number of support points, PSO appears to be far superior to the other algorithms for searching for $D$-optimal designs for GLMs with all continuous factors and a binary response.

\begin{table}
\caption{\label{cont_speed_sims} Simulation results (in seconds of CPU time) to find 1000 locally $D$-optimal designs for a 2 continuous factor model at different resolutions. }
\centering
\begin{tabular}{|c|ccc|}
\hline
Resolution & \multicolumn{1}{c|}{Lift-One} & \multicolumn{1}{c|}{Cocktail} & Fedorov-Wynn \\ \hline
0.10       & 1046.63                       & 37.68                         & 1754.73      \\
0.05       & NA                            & 109.27                        & 6435.06      \\
0.01       & NA                            & 2810.77                       & NA           \\ \hline
\multicolumn{4}{|l|}{PSO required only 30.41 seconds to find the 1000 designs }      \\ \hline
\end{tabular}
\end{table}

Yang et al. (2011) developed a complete class approach to finding optimal designs when all factors are continuous. For any design that is not a member of the complete class they identified a design in their complete class that dominates another under the Loewner ordering. If there are $k$ factors in the model, designs that belong to their complete class are supported at the boundaries of $k-1$ factors and are unrestricted for the unbounded factor. Further, designs in this class have at most $2^k$ support points. A disadvantage of this approach is that it requires one factor to be unbounded because in practice all factors are bounded due to physical constraints. If practical constraints require a truncation of the unbounded range of the last factor, designs calculated from the theory might not be efficient or implementable.

Consider the design problem given in Stufken and Yang (2011), where all three factors are continuous with $x_1 \in [-2,2]$,  $x_2 \in [-1,1]$, and $x_3 \in (-\infty, \infty)$. The model of interest is $logit(\mu) = \beta_0 + \beta_1 x_1 + \beta_2 x_2 + \beta_3 x_3$ and the nominal values are: $(\beta_0,\beta_1,\beta_2,\beta_3) = (1, -0.5, 0.5, 1)$. Without restricting $x_3$, Stufken and Yang (2011) produced the locally $D$-optimal design equally supported at eight points shown in the left panel in Table~\ref{continuousdesigns}. The result in Yang et al. (2011) does not rule out the possibility of having $D$-optimal designs with fewer support points, but it is unable to identify them. PSO was able to find the equally efficient design with only four support points shown in the middle panel of Table~\ref{continuousdesigns}. This design was found by PSO using 25 particles, convergence tolerance 0.0001, 150 maximum iterations, 100 maximum resets, and a minimum lower efficiency bound of 99\%. We observe that the PSO-generated design requires half as many design points and that they are a proper subset of the support points identified in the design found by Stufken and Yang (2011). Designs with fewer support points are in general more cost-efficient and hence desirable.

\begin{table}
 \caption{\label{continuousdesigns} The left panel shows the design found by the theory in Yang et al. (2011). The middle panel is the PSO-generated $D$-optimal design with a truncated interval $[-10, 10]$ for the third factor. The right hand panel is the design found by PSO when the third factor was constrained further to $[-2,2]$.   \vspace{0.2in}
}
  \centering
  \makebox{
  \begin{tabular}{|ccc|c|}
  \hline
$\phantom{-}X_1$ & $\phantom{-}X_2$ & $\phantom{-}X_3$ & $p_i$\\
$-2$ & $-1$ & $-0.456$ & $0.125$\\
$-2$ & $-1$ & $-2.544$ & $0.125$\\
$-2$ & $\phantom{-}1$ & $-1.456$ & $0.125$\\
$-2$ & $\phantom{-}1$ & $-3.544$ & $0.125$\\
$\phantom{-} 2$ & $-1$ & $\phantom{-}1.544$ & $0.125$\\
$\phantom{-}2$ & $-1$ & $-0.544$ & $0.125$\\
$\phantom{-}2$ & $\phantom{-}1$ & $\phantom{-}0.544$ & $0.125$\\
$\phantom{-}2$ & $\phantom{-}1$ & $-1.544$ & $0.125$\\
  \hline
  \end{tabular}
  }
  \makebox{
  \begin{tabular}{|ccc|c|}
  \hline
$\phantom{-}X_1$ & $\phantom{-}X_2$ & $\phantom{-}X_3$ & $p_i$\\
$-2$ & $-1$ & $-2.544$ & $0.25$\\
$-2$ & $\phantom{-}1$ & $-1.457$ & $0.25$\\
$\phantom{-}2$ & $-1$ & $\phantom{-}1.544$ & $0.25$\\
$\phantom{-}2$ & $\phantom{-}1$ & $-1.544$ & $0.25$\\
  \hline
  \end{tabular}
  }
  \makebox{
  \begin{tabular}{|ccc|c|}
  \hline
$\phantom{-}X_1$ & $\phantom{-}X_2$ & $\phantom{-}X_3$ & $p_i$\\
$-2$ & $-1$ & $-2.000$ & $0.212$\\
$-2$ & $\phantom{-}1$ & $-2.000$ & $0.043$\\
$-2$ & $\phantom{-}1$ & $-1.649$ & $0.166$\\
$\phantom{-}2$ & $-1$ & $\phantom{-}1.745$ & $0.214$\\
$\phantom{-}2$ & $-1$ & $-0.748$ & $0.075$\\
$\phantom{-}2$ & $\phantom{-}1$ & $-1.748$ & $0.214$\\
$\phantom{-}2$ & $\phantom{-}1$ & $\phantom{-}0.748$ & $0.075$\\
  \hline
  \end{tabular}
  }
\end{table}

\begin{figure}[H]
\begin{center}
\begin{tabular}{c}
\vspace{-.4in}\includegraphics[width=2.5in]{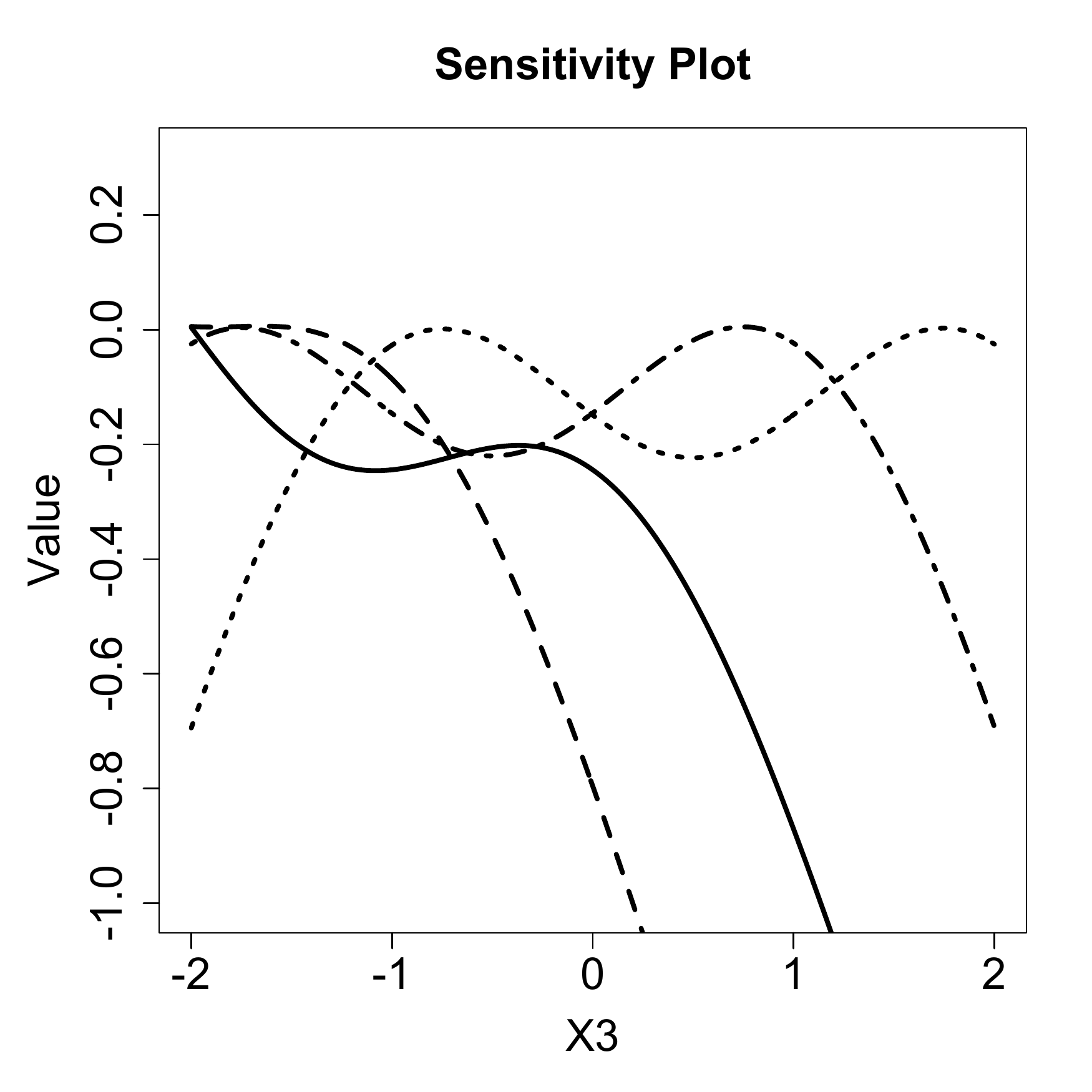}\\
\vspace{-2.25in}\includegraphics[width=3in]{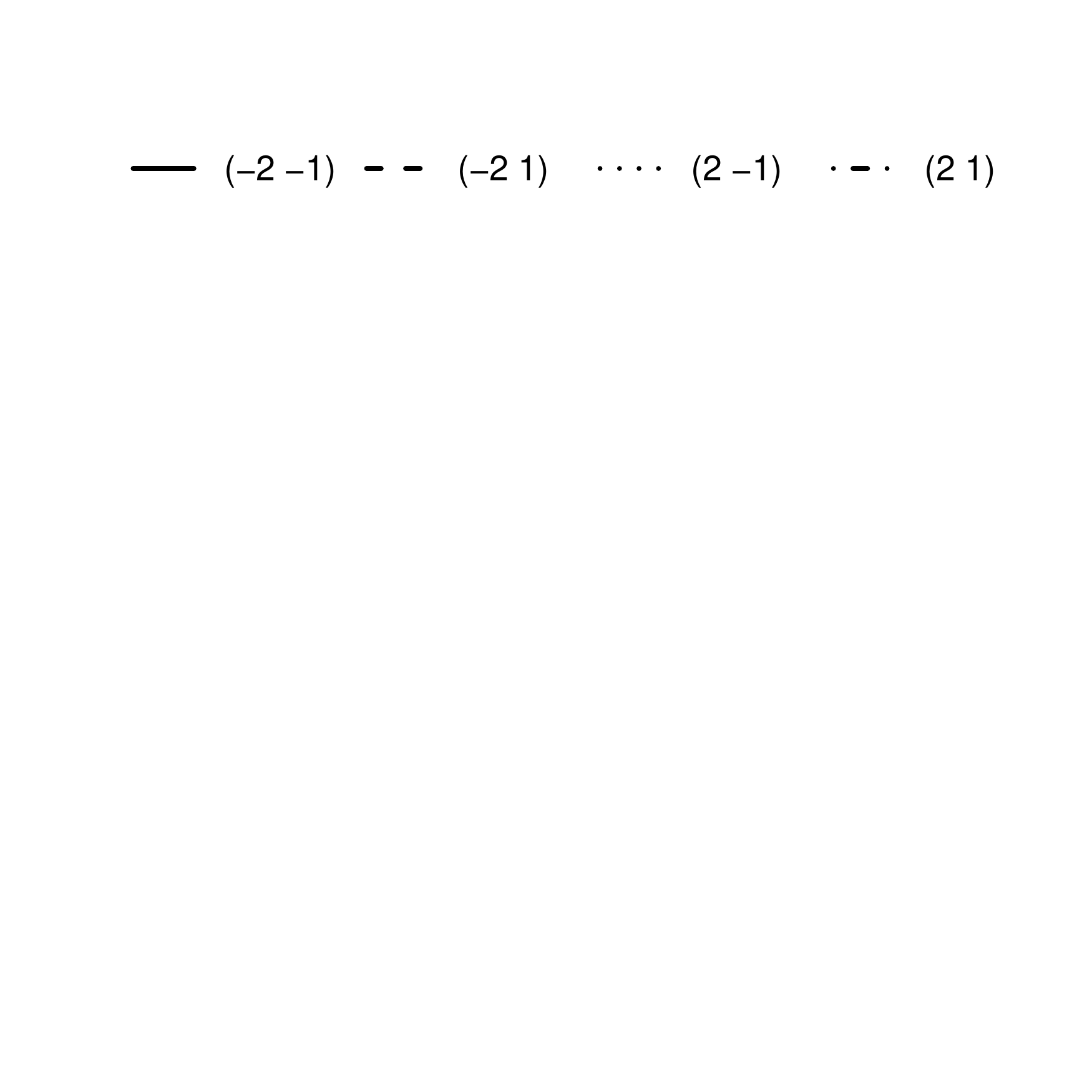}
\end{tabular}
\caption{The sensitivity plot for the restricted experiment with three continuous factors.}
\label{equiv_restricted}
\end{center}
\end{figure}

We notice that two of the support points of the two designs have values larger than 2. What if the experimenter needs to limit the values of $x_3$ to the interval $[-2,2]$? PSO finds the locally $D$-optimal design on the smaller interval easily using the same settings, and the unequally weighted seven point design is show in the right hand panel of Table~\ref{continuousdesigns}. Figure~\ref{equiv_restricted} shows the sensitivity plot of this design and confirms its optimality.

\subsection{Discrete and Continuous Factors}
In this subsection, we use PSO to find locally $D$-optimal designs for logit models with a couple of mixed factors and delineate cases when and if a minimally supported optimal design exists. As a start, we evaluate PSO's ability to find designs with both discrete and continuous factors by simulating the simplest case in which we have one discrete factor and one continuous factor. The model is $logit(\mu) = \beta_0 + \beta_1 x_1 + \beta_2 x_2$ where $x_1 \in \{-1,1\}$ and $x_2 \in [-1,1]$ and the range of values for the nominal values  are $\beta_0 \in \{1, 1.5, 2\}$, $\beta_1 \in [-1.5, 1.5]$ and $ \beta_2 \in [-3,3].$ We employ PSO to find locally $D$-optimal designs using 25 particles, up to 100 maximum iterations and up to 500 maximum resets.  The convergence tolerance is 0.0001 and the  minimum lower efficiency bound used to find the locally optimal design is set at 99.9\%. For the simulation, we discretize the  parameter space for $\beta_1$ and $\beta_2$ using a grid with resolution 0.01 and construct designs for all combinations of $\beta_1, \beta_2$. This results in a total of 180,901 (301 $\beta_1$ values $\times$ 601 $\beta_2$ values) locally $D$-optimal designs for each $\beta_0$ level.

\begin{figure}[H]
\begin{center}
\includegraphics[width=2.4in]{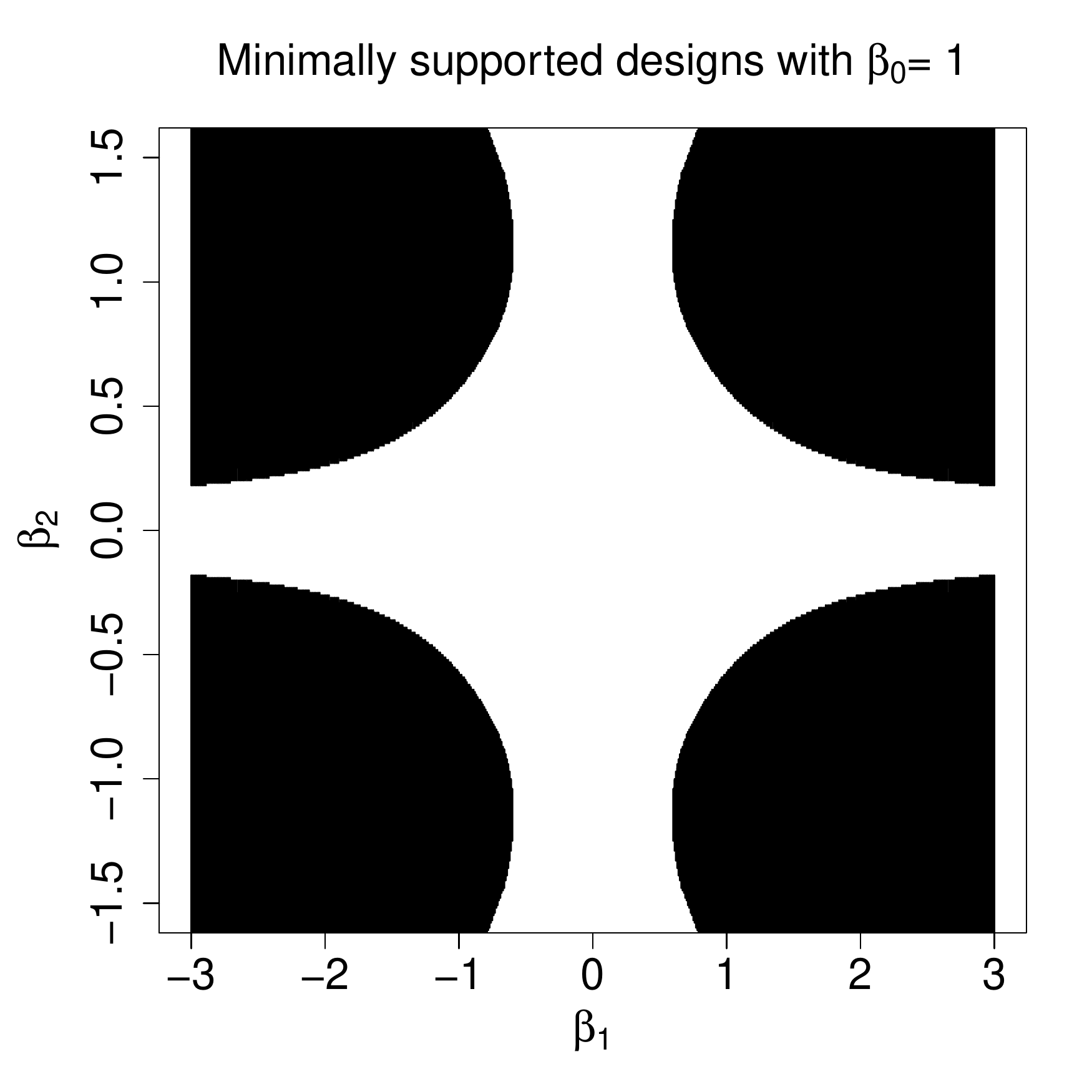}
\includegraphics[width=2.4in]{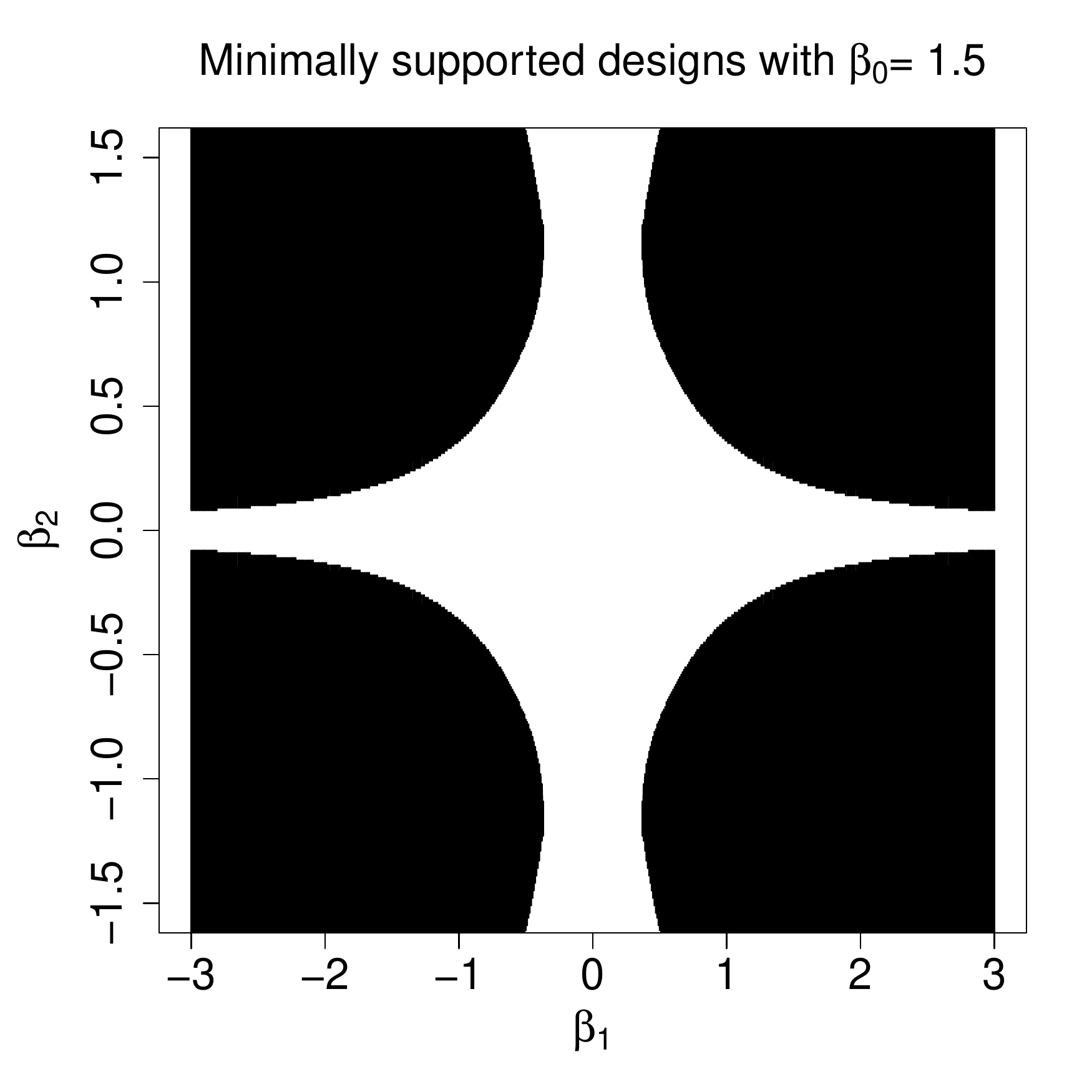}
\includegraphics[width=2.4in]{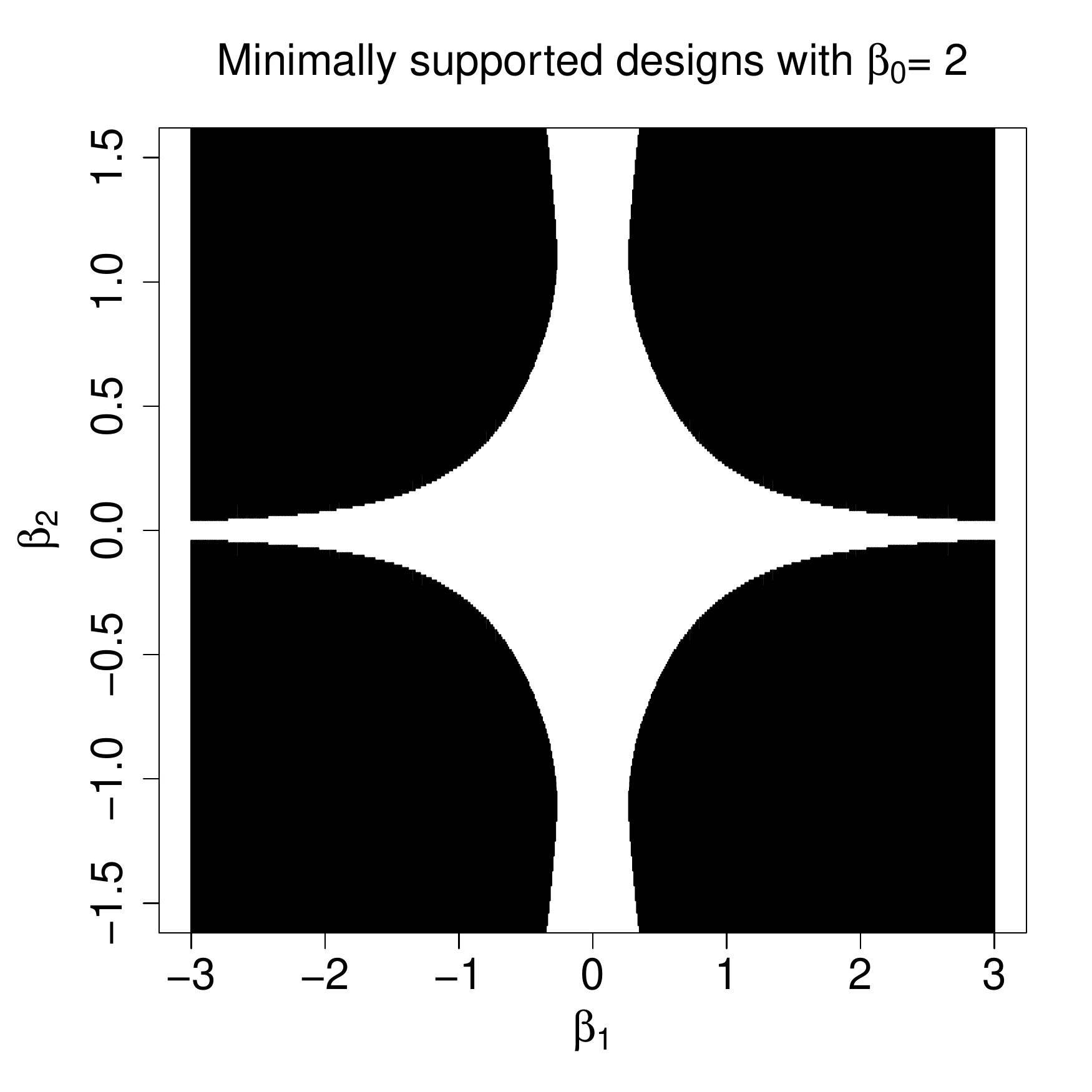}
\includegraphics[width=2.4in]{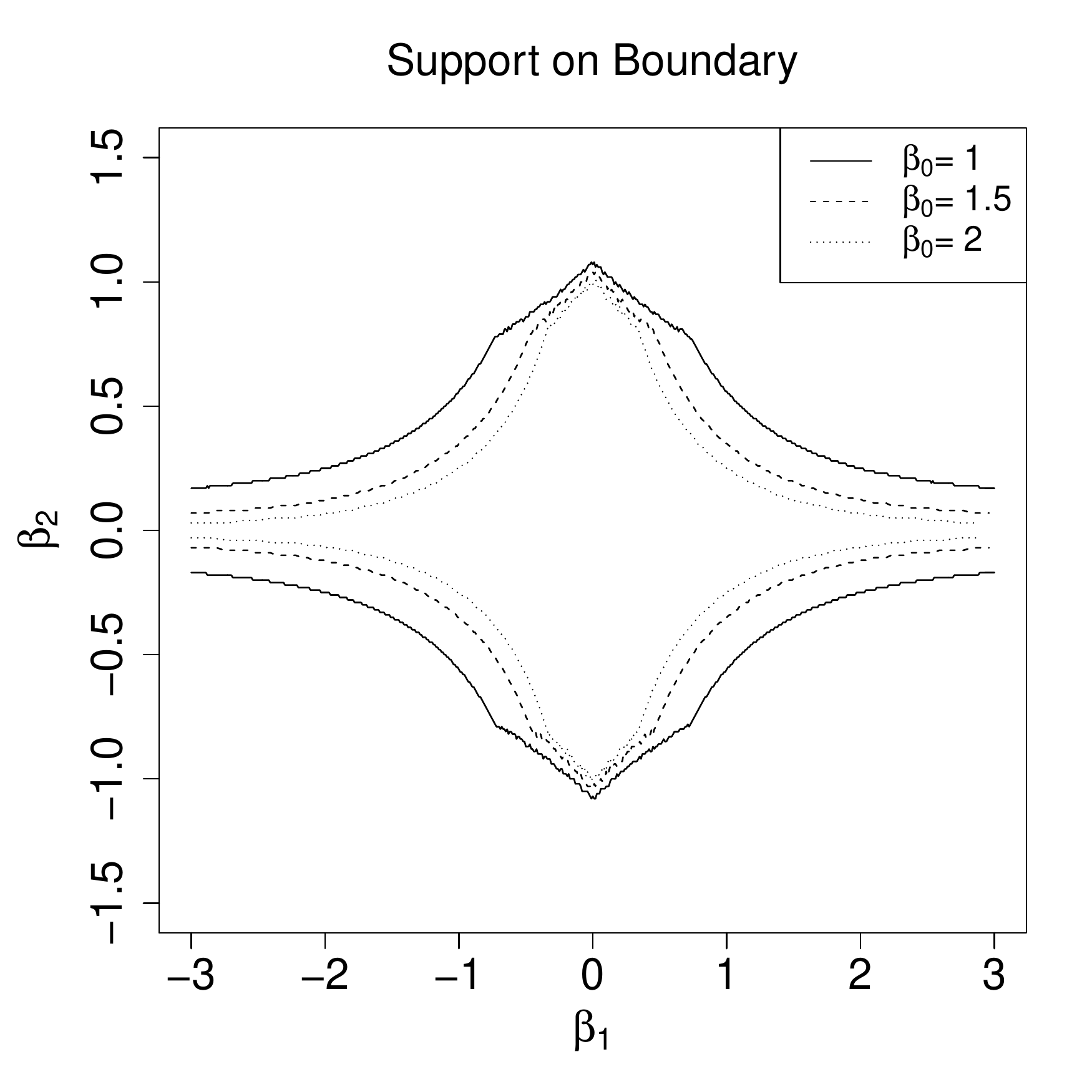}
\caption{The first 3 panels contains plots of where minimally supported designs can be constructed with $\beta_0 = {1, 1.5, 2}$. Areas in which minimally supported designs can be constructed are in black. The final panel displays the region within which designs can be constructed using points only on the boundary.}
\label{disconsuppts}
\end{center}
\end{figure}

The black curvilinear areas in the first three panels of Figure~\ref{disconsuppts} show parameter values for $\beta_1$ and $\beta_2$ for which minimally supported designs could be constructed when $\beta_0 = \{1, 1.5, 2\}$. Similar figures can be generated for other nominal values for a fixed value of $\beta_0$. As the magnitude of $\beta_0$ increases, the region in which a minimally supported design can be constructed increases as well.  The final panel in the same figure displays the values for $\beta_1$ and $\beta_2$  for which non-minimally supported designs can be constructed using only points on the boundary when $\beta_0=0$. The lines correspond to the boundaries within which 4 point designs can be constructed with $x_2 \in \{-1,1\}$. The figure shows that as the magnitude of the intercept increases fewer designs can be constructed using only boundary points.

\subsubsection{Robustness Under Model Mis-specification}
Before a design is implemented, it is important to investigate its robustness properties to model mis-specification. For example, in binary response GLMs,  a design constructed under the probit link may perform poorly if the true link function is the complementary log-log. In practice it is common to choose the logit link, but a prudent researcher should choose a design that reflects the actual goals of the study and has acceptable efficiency if there are violations in the model assumptions.  Invariably, for many real problems there are competing objectives in the study and different degrees of concern on various aspects of model adequacy. Here we focus on the issue of model misspecification, but other aspects of violations of the GLM assumptions can be studied in a similar manner. Ideally, the user should carefully appreciate the efficiency differences of a design under various criteria and model mis-specifications  before implementing it. In what follows, we use PSO to investigate the robustness of the locally $D$-optimal designs found under  the logit link when the true link function is probit, log-log, or complementary log-log.

We run PSO with 25 particles and terminate the algorithm if the design found is at least 99\% efficient or if the maximum number of resets is reached. The maximum number of iterations allowed is 100 with 500 maximum resets. We compare the relative efficiency of the logit link based PSO-generated design to PSO-generated designs under the correct link function. The swarm terminates its search if the equivalence theorem is satisfied or if the maximum number of swarm resets is reached. In this study we considered the model $logit(\mu) = \beta_0 + \beta_1 x_1 + \beta_2 x_2$ with $x_1 \in \{-1,1\}$ and $x_2 \in [-1,1]$, but other models can be used as well. We take $\beta_0 = 1$ and explored $\beta_1 \in [-1.5,1.5]$ and $\beta_2 \in [-3, 3]$ over a grid with resolution 0.1. This means that each dimension of the search space is discretized using equally spaced grid points 0.1 apart; for our case, this results in a total of 1,891 individual optimal designs built around each link function. We then compare the relative efficiency of the logit model to each of the probit, log-log, and complementary log-log designs under the assumption that the probit, log-log, and complementary log-log link function was the true link function and our choice of the logit link was incorrect.

\begin{figure}[]
\begin{center}
\includegraphics[width=2.1in]{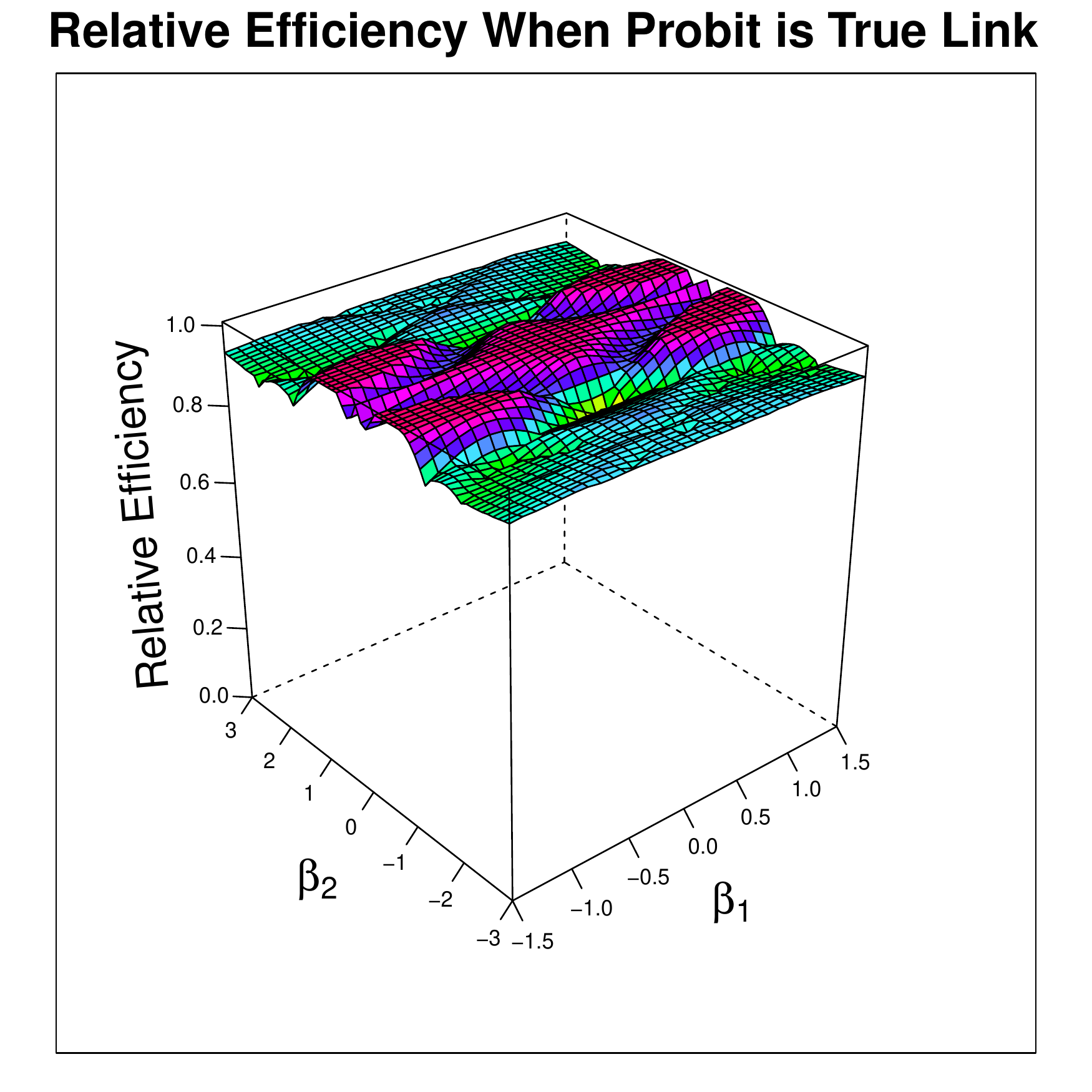}
\includegraphics[width=2.1in]{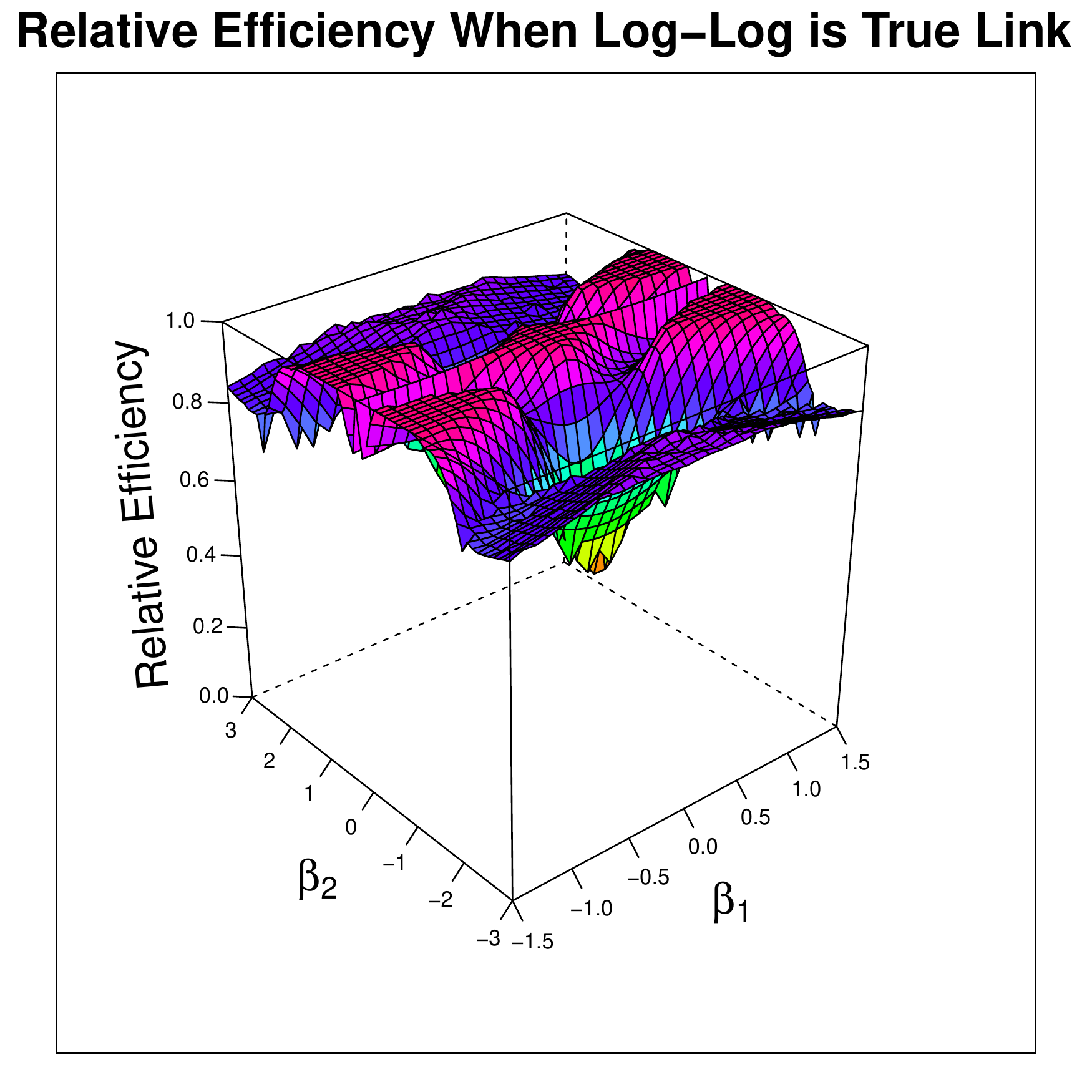}
\includegraphics[width=2.1in]{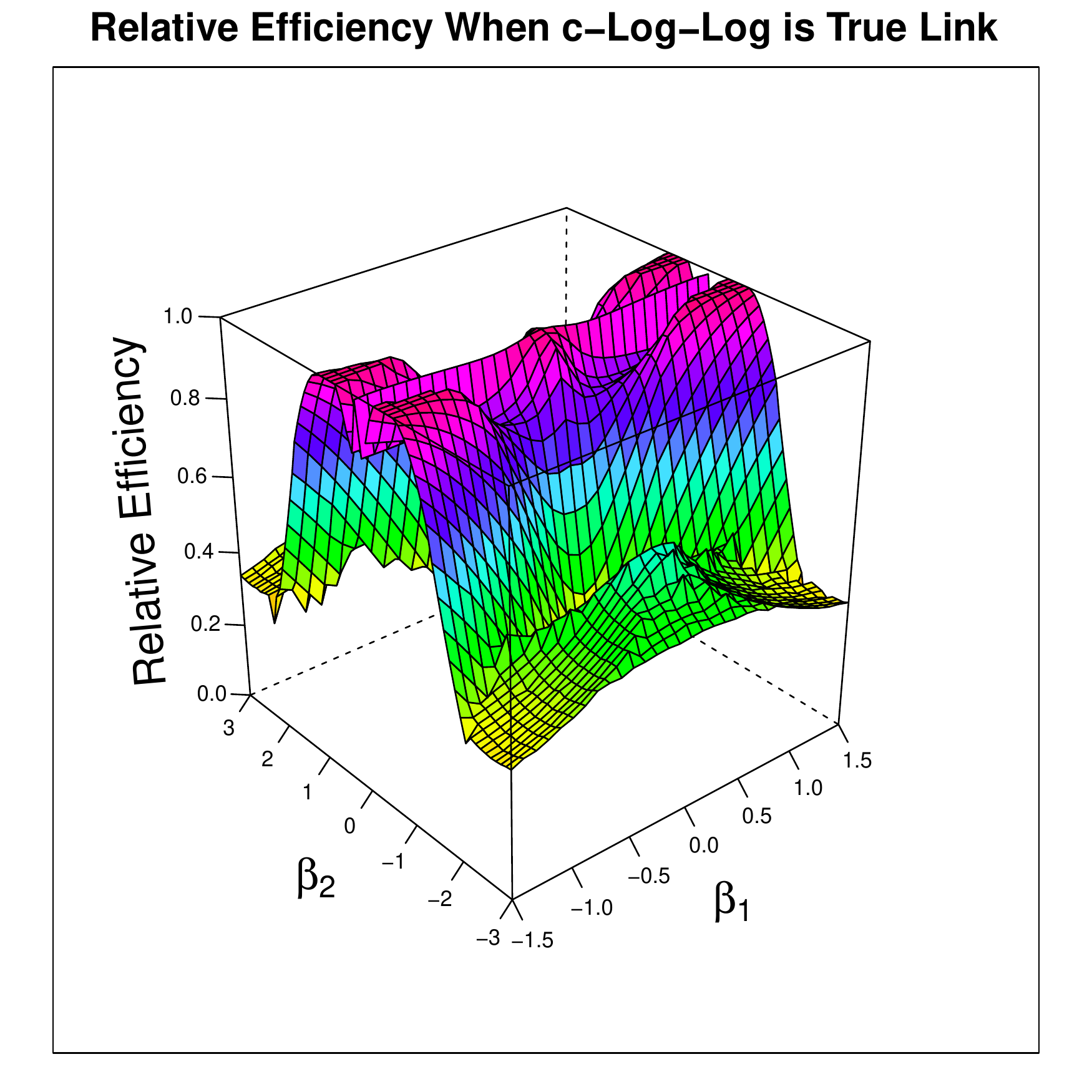}
\caption{Relative efficiency of the logit link based locally $D$-optimal design to the probit, log-log, and complementary log-log link based optimal designs.}
\label{modelmisgraphs}
\end{center}
\end{figure}

Figure \ref{modelmisgraphs} and Table \ref{modelmistable} provide the results of these simulations. From the table we can see that the majority of the designs obtained by PSO are fairly robust against model mis-specification. When the true link is the probit or log-log link, the logit-based designs work quite well. When the complementary-log-log link is the true link function the designs obtained by PSO tend to perform somewhat more poorly. From a visual inspection of the relative efficiencies in Figure \ref{modelmisgraphs} we observe that while the logit-based designs are generally robust, the designs can perform extremely poorly when the true link is the log-log or the complementary-log-log. The problematic locations appear to occur when the effect of $\beta_1$ is near 0 for the log-log link and when the effects of $\beta_1$ and $\beta_2$ are both near their extremes for the complementary log-log link. Finally, note that these results are all for $\beta_0 = 1$. For different values of $\beta_0$ the efficiencies of designs with the incorrect link function may behave differently.

\begin{table}
\caption{\label{modelmistable} Quantiles of the relative efficiencies of the logit link based designs when the true link functions are probit, log-log, and complementary-log-log.}
\centering
\begin{tabular}{|c|ccc|}
\hline
     \diagbox{Quantile}{True Link}                   & Probit                     & Log-log                    & C-log-log                  \\
\hline
0.99                    & 1.0000                     & 1.0000                     & 1.0000                     \\
0.95                    & 1.0000                     & 1.0000                     & 0.9900                     \\
0.90                     & 1.0000                     & 1.0000                     & 0.9488                     \\
0.80 & 0.9900 & 0.9737 & 0.8692 \\
0.70 & 0.9670 & 0.9106 & 0.7925\\
\hline
\end{tabular}
\end{table}

\section{Applications}

We revisit the motivating application at the start of the paper and show the PSO-generated $D$-optimal design can be much more efficient than the design implemented by the researchers.  As another application, we also apply PSO to find an efficient design for an electrostatic discharge study.

\subsection*{Application 1: A revisit to the Odor Removal Study}

\begin{table}
 \caption{\label{odorcompare} Relative efficiencies of factorial designs for the odor experiment to PSO's design when breaking up temperature at different intervals. \vspace{0.2in}}
  \centering
  \makebox{
  \begin{tabular}{|c|ccccc|}
  \hline
  & 1 Unit & 3 Units & 5 Units & 10 Units & 15 Units \\
  \hline
  Support Points & 496 & 176 & 112 & 64 & 48 \\
  Relative Efficiency & 0.5610 & 0.5675 & 0.5730 & 0.5831 & 0.5896 \\
  \hline
  \end{tabular}
  }
\end{table}

\begin{figure}
\centering

\begin{tabular}{cc}

\begin{minipage}{0.46\textwidth}
\begin{tabular}{l}
\vspace{-0.1in}\includegraphics[width=3in]{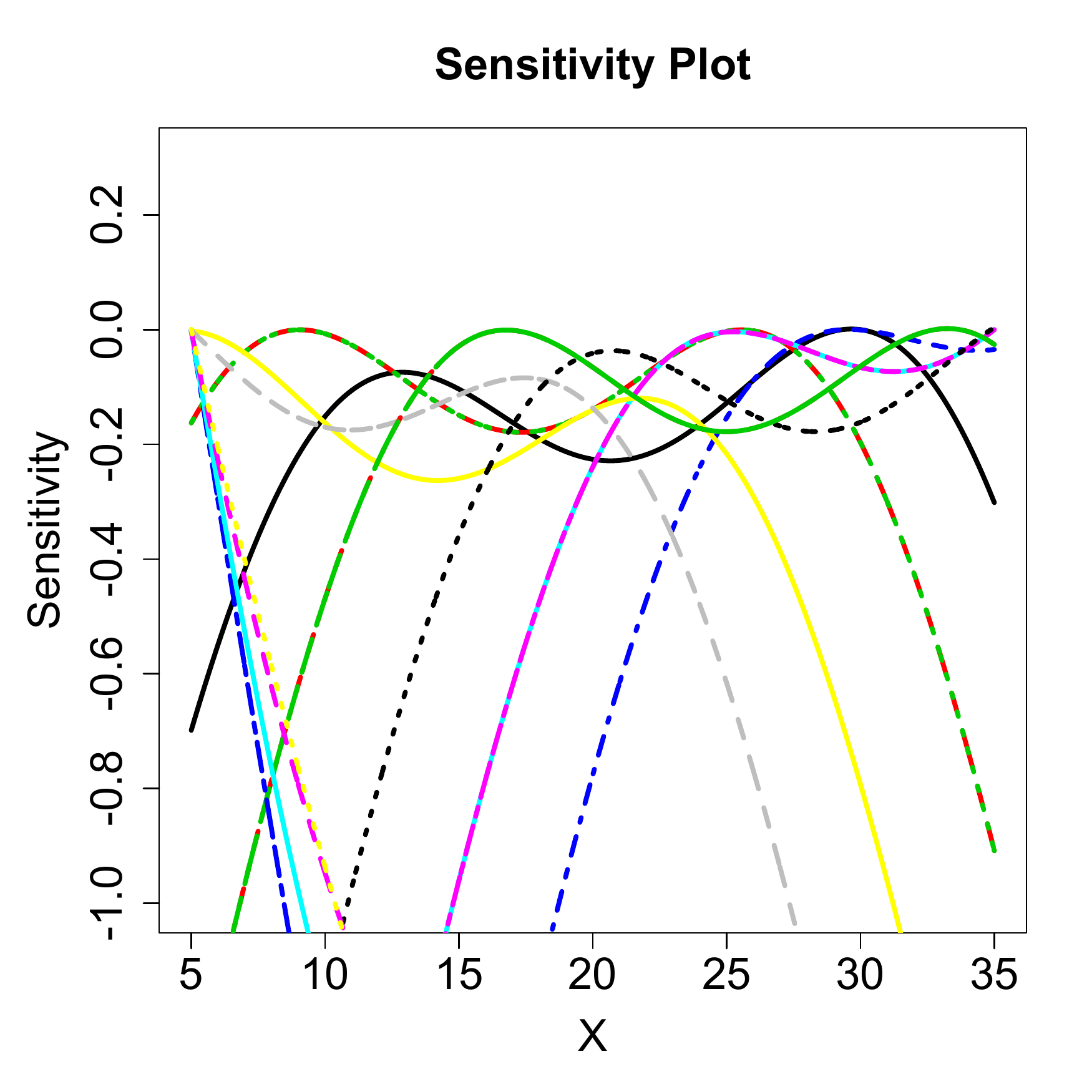}\\
\hline \vspace{-2in}
\vspace{-1.3in}\includegraphics[width=3in]{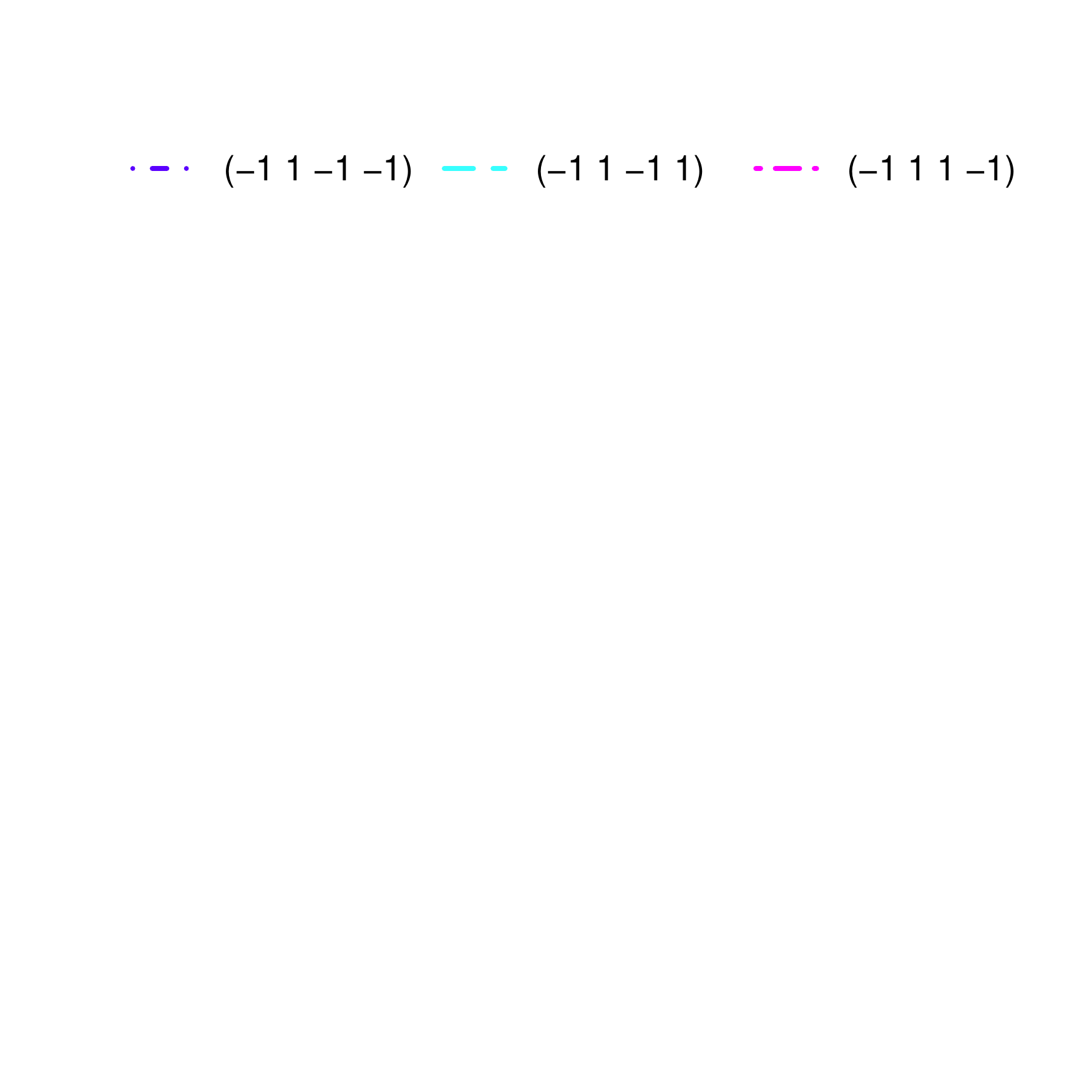}\\
\vspace{-2.65in}\includegraphics[width=3in]{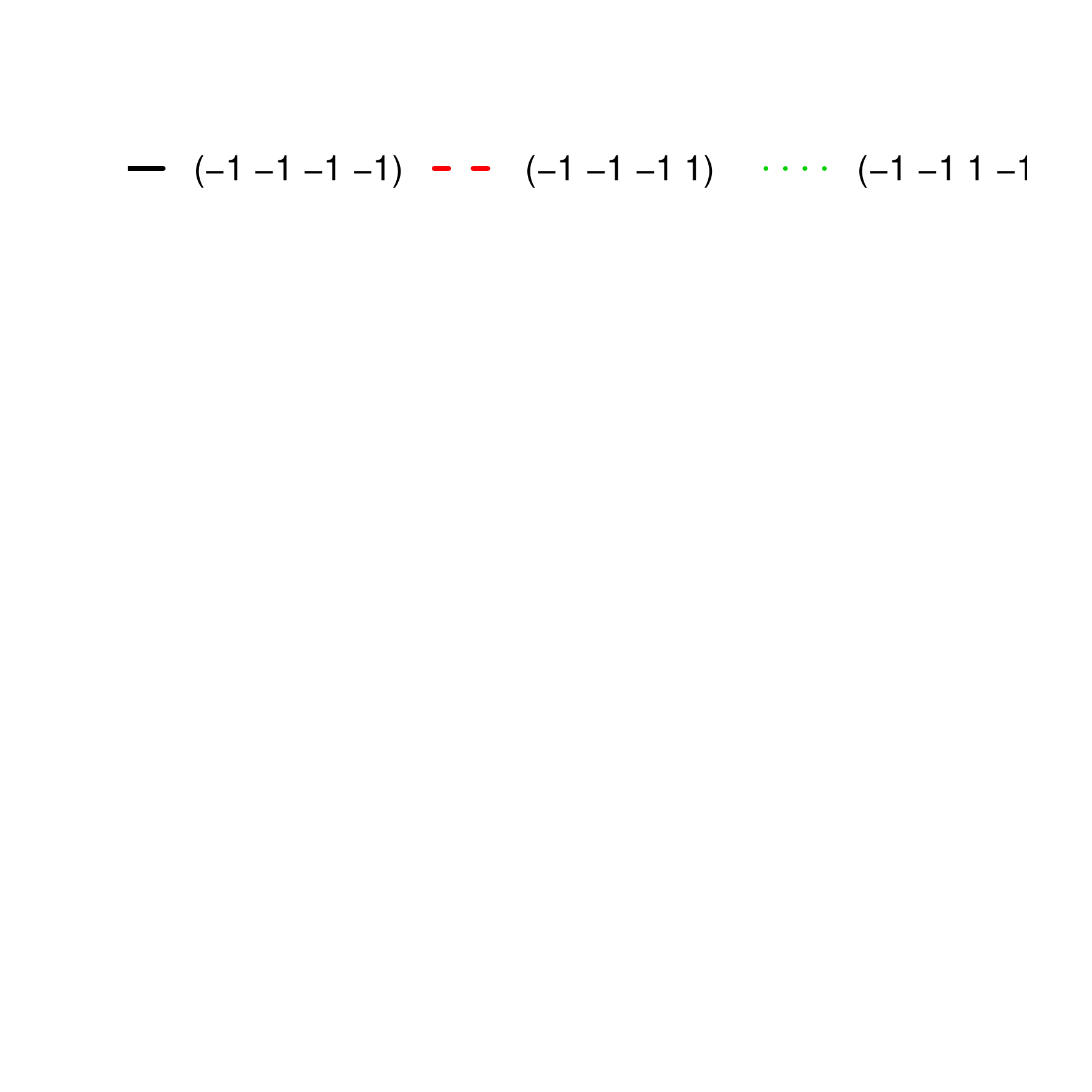}\\
\vspace{-2.85in}\includegraphics[width=3in]{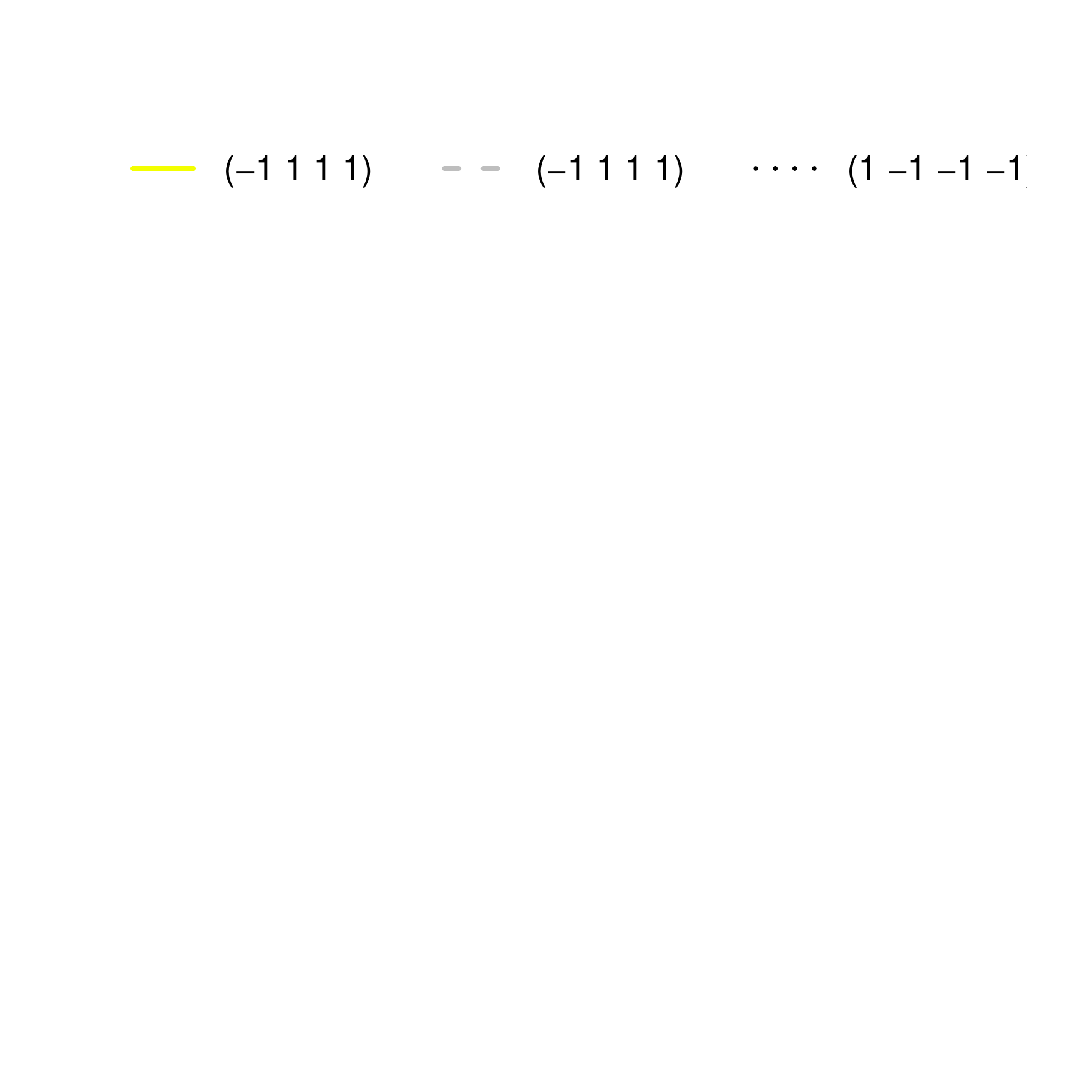}\\
\vspace{-2.4in}\includegraphics[width=3in]{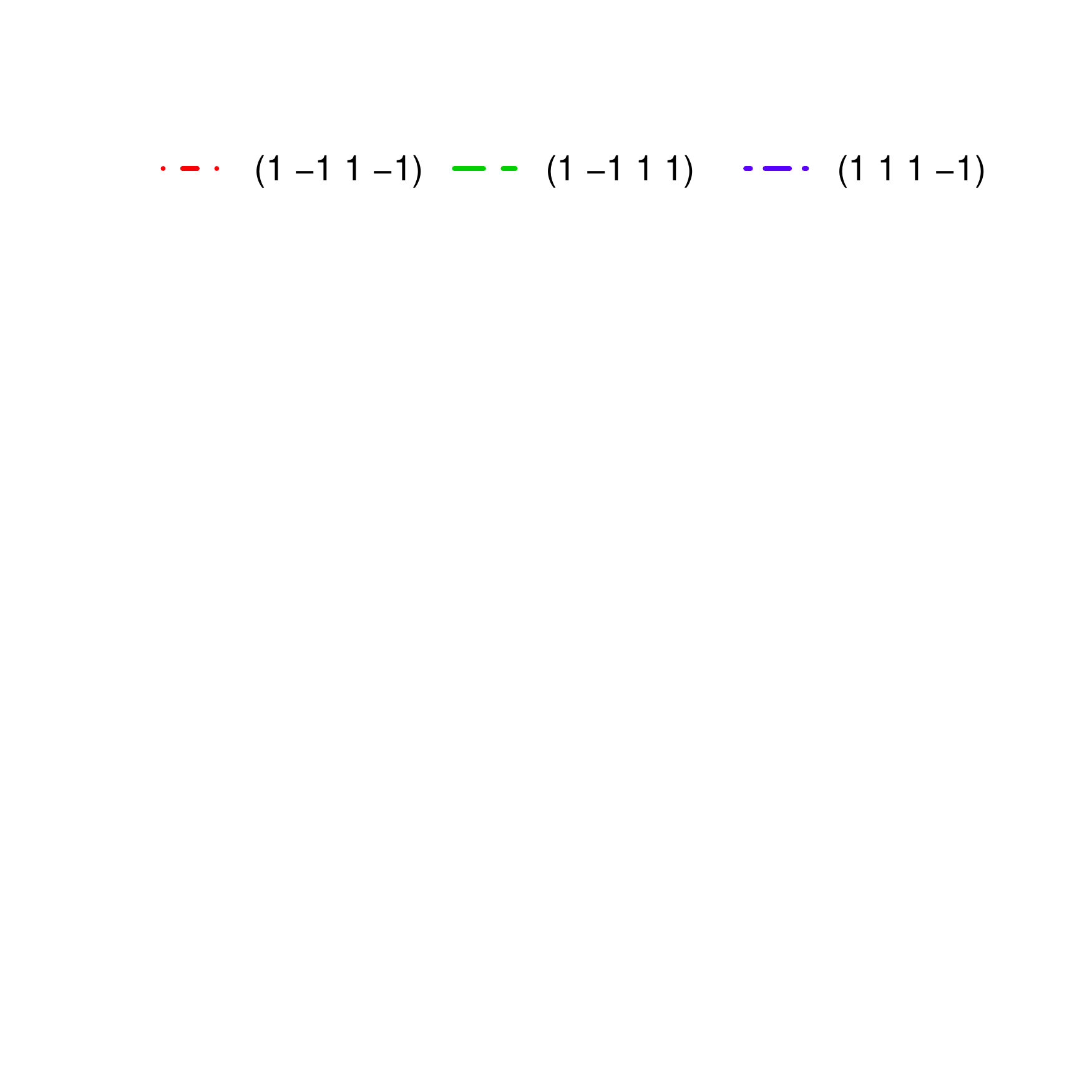}\\
\hline
\end{tabular}
\end{minipage}

&

\begin{minipage}{0.6\textwidth}
  \begin{tabular}{|ccccc|c|}
  \hline
Algae & Scav. & Resin & Comp. & Temp. & $p_i$\\
$-1$ & $-1$ & $-1$ & $-1$ & 29.67 & 0.1022\\
$-1$ & $-1$ & $-1$ & $\phantom{-}1$ & 25.56 & 0.0440\\
$-1$ & $-1$ & $-1$ &  $\phantom{-}1$ & 9.082 & 0.0339 \\
$-1$ & $-1$ & $\phantom{-}1$ & $\phantom{-}1$ & 29.58 & 0.1143 \\
$-1$ & $-1$ & $\phantom{-}1$ &$\phantom{-}1$ & 25.26 & 0.0052\\
$-1$ & $-1$ &  $\phantom{-}1$ & $\phantom{-}1$ & 35.00 & 0.0445\\
$-1$ & $\phantom{-}1$ & $-1$ & $-1$ & 5.000 & 0.0793 \\
$-1$ & $\phantom{-}1$ & $-1$ & $\phantom{-}1$ & 5.000 & 0.0989\\
$-1$ & $\phantom{-}1$ & $\phantom{-}1$ & $-1$ & 35.00 & 0.0618 \\
$-1$ & $\phantom{-}1$ & $\phantom{-}1$& $\phantom{-}1$ & 33.22 & 0.0891\\
$-1$ &  $\phantom{-}1$ & $\phantom{-}1$ & $\phantom{-}1$ & 16.76 & 0.0189\\
$\phantom{-}1$ &  $-1$ &  $-1$ &  $-1$ & 5.000 & 0.0511\\
$\phantom{-}1$ & $-1$ & $\phantom{-}1$ & $-1$ & 5.000 & 0.0526\\
$\phantom{-}1$ & $-1$ & $\phantom{-}1$ & $\phantom{-}1$ & 5.000 & 0.1072\\
$\phantom{-}1$ & $\phantom{-}1$ &  $\phantom{-}1$ & $-1$ & 5.000 & 0.0968\\
\hline
  \end{tabular}
\end{minipage}

\end{tabular}
\caption{The sensitivity plot for the odor experiment. \textbf{Table 7.} The design for the odor experiment obtained using PSO. }
\label{odor_equiv}
\end{figure}

We now return to the odor removal study conducted by Wang et al. (2015). We recall the experimenters chose not to test the effect of temperature because they were not able to construct a theoretical optimal design capable of accommodating the mixed factors. They ignored the temperature variable and used a $2^{4-1}$ regular fractional factorial design with equal number of replicates. The reported parameter estimates were $2.89$, $0.84$, $-1.47$, and $-0.02$ for the effects of algae type, scavenger type, resin, and compatibilizer respectively. PSO allows us to construct a design that includes storage temperature as a continuous factor for the model $logit(\mu) = \beta_0 + \beta_1 Algae + \beta_2 Scavenger + \beta_3 Resin + \beta_4 Compatibilizer + \beta_5 Temperature$. Our nominal values were $\mathbf{\hat{\beta}} = (-1, 2, 0.5, -1, 0.25, 0.13)$, where the nominal values for the intercept and temperature were based on their expected effects. The parameters for PSO were 100 particles, 400 maximum iterations, a convergence tolerance of 0.0001, 1000 maximum resets, and a minimum $D$-efficiency lower bound of 99\% to identify a $D$-optimal design for this experiment. The PSO-generated design is given in Table 7 and its sensitivity plot in Figure \ref{odor_equiv} confirms its $D$-optimality.

The PSO-generated design has 15 unique support points, with only some of them at the boundaries of the continuous factor. An experiment obtained by discretizing the temperature factor would likely not have included these points, and so it is likely that the resulting design would be less efficient. To gain insight, we break up temperature along its range using several different resolutions and calculate the $D$-efficiencies of the factorial designs using the discretized versions of temperature. Table \ref{odorcompare} displays the relative efficiencies and the number of support points of these designs. We observe that designs obtained this way have many more support points than the PSO-generated design and are only about half as efficient.


\addtocounter{table}{1}


\subsection*{Application 2: Electrostatic Discharge Study }
A similar experiment  with mixed continuous and discrete factors is described in Whitman et al. (2006). The experimenters were interested in finding what factors influence the failure of semiconductors when exposed to electrostatic discharge (ESD). The response was whether or not a certain part of the semiconductor failed, and the model was  logistic regression with  5 factors shown in Table~\ref{voltagefactors}. The first two factors, Lot A and Lot B, were used to describe the type of wafer used. ESD testing requires a part to be ``zapped'' with a pulse at either a positive or a negative polarity and then to be zapped again by a pulse with the opposite polarity. A lack of standardization of which pulse order should be used resulted in the experimenters using pulse order as the third factor. The fourth factor was ESD handling: whether or not proper ESD precautions were taken. The final factor was continuous: the voltage at which the chip was tested. The experimental units were a single chip TDMA power amplifiers, chosen for their lack of ESD protection circuitry. Without protection circuitry the chips were expected to be sensitive to changes in experimental factors. The model of interest was $logit(\mu) = \beta_0 + \beta_1 Lot A + \beta_2 Lot B + \beta_3 ESD + \beta_4 Pulse + \beta_5 Voltage + \beta_{34}(ESD \times Pulse)$.

The experimenters treated the voltage variable as a discrete factor with 5 levels: 25, 30, 35, 40, and 45 volts so that all factors are now discrete. The full factorial design with 80 separate settings was implemented to test all possible combinations of factor settings. They did not provide a rationale for not using an experiment with fewer runs but it is likely because there was no known method for optimally designing such a study with mixed factors at that time.

\begin{table}
\caption{\label{voltagefactors} Factor types and levels for the electrostatic discharge experiment. \vspace{0.2in}}
  \centering
  \makebox{
\begin{tabular}{|c|c|cc|}
\hline
Type      & Factor  & \multicolumn{2}{c|}{Levels} \\
&         &  $-$ & $+$ \\
\hline
\multirow{4}{*}{Discrete}         & Wafer Lot A   & Position 1 &  Position 2\\
 & Wafer Lot B    &  Position 1 &  Position 2\\
& ESD      & No & Yes \\
&  Pulse   & negative & positive\\
\hline
Continuous & Voltage & \multicolumn{2}{c|}{Voltage from 25 to 45}\\
\hline
\end{tabular}
  }
\end{table}

\begin{figure}
\centering

\begin{tabular}{cc}

\begin{minipage}{0.5\textwidth}
\begin{tabular}{l}
\vspace{-0.1in}\includegraphics[width=3in]{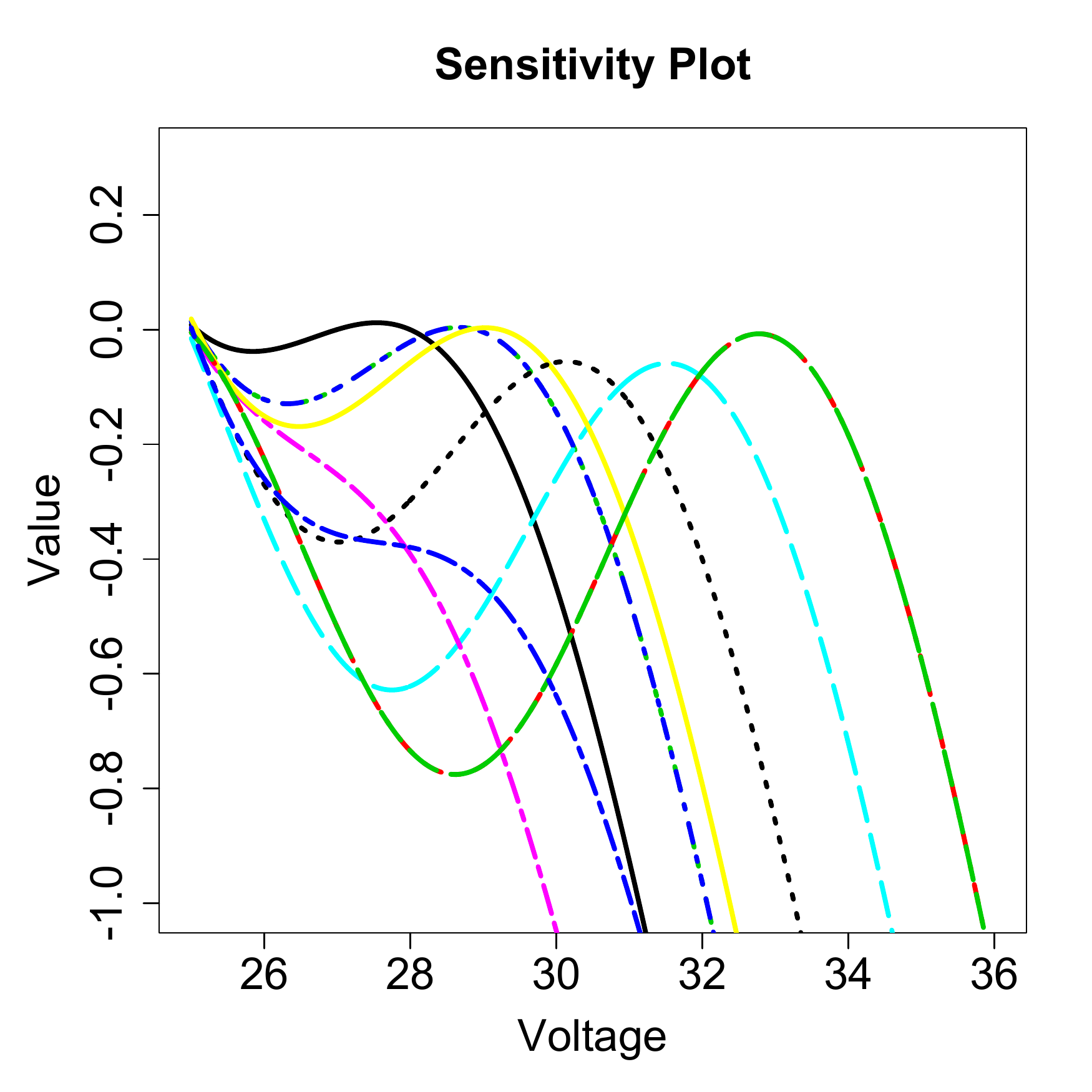}\\
\hline \vspace{-2in}
\vspace{-1.3in}\includegraphics[width=3in]{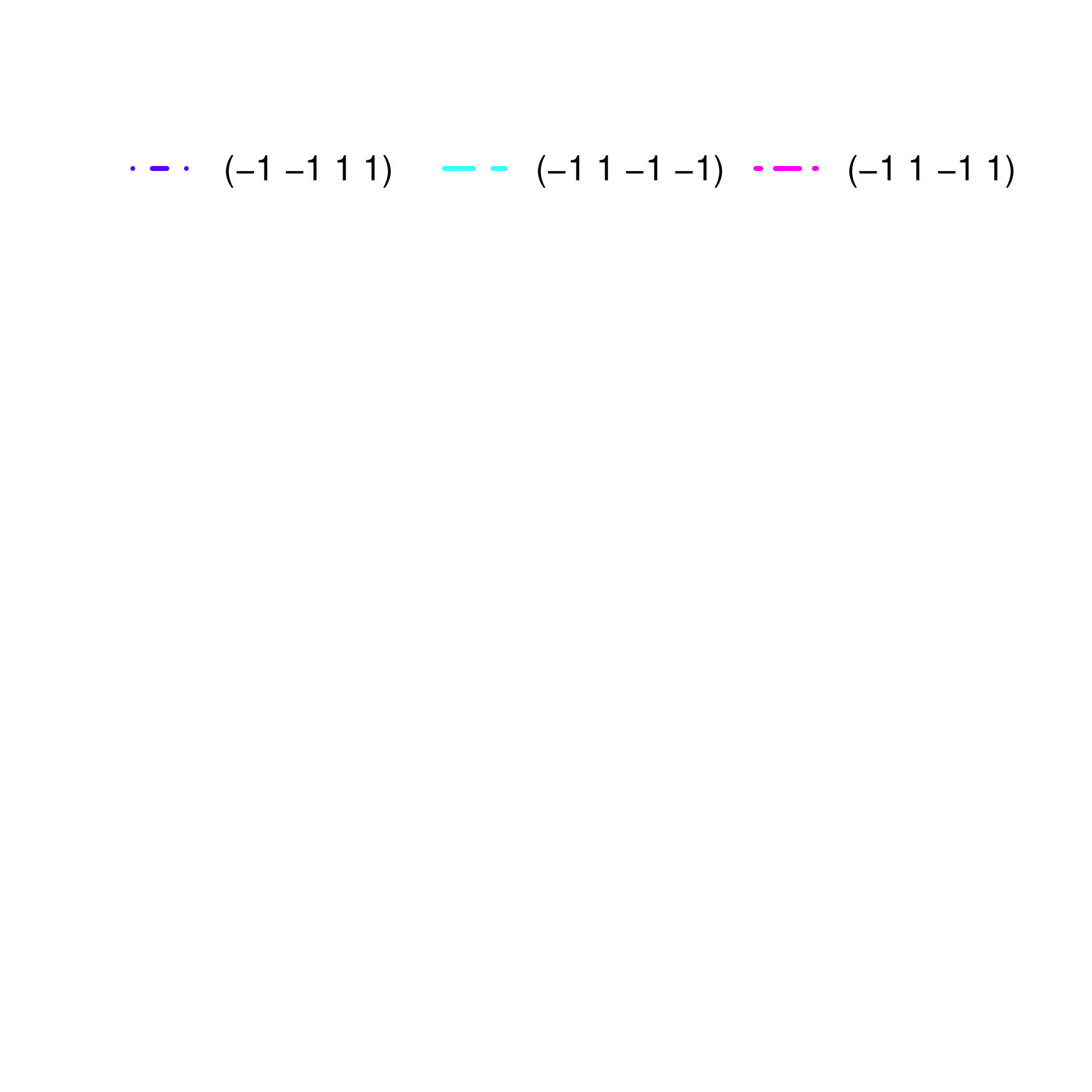}\\
\vspace{-2.65in}\includegraphics[width=3in]{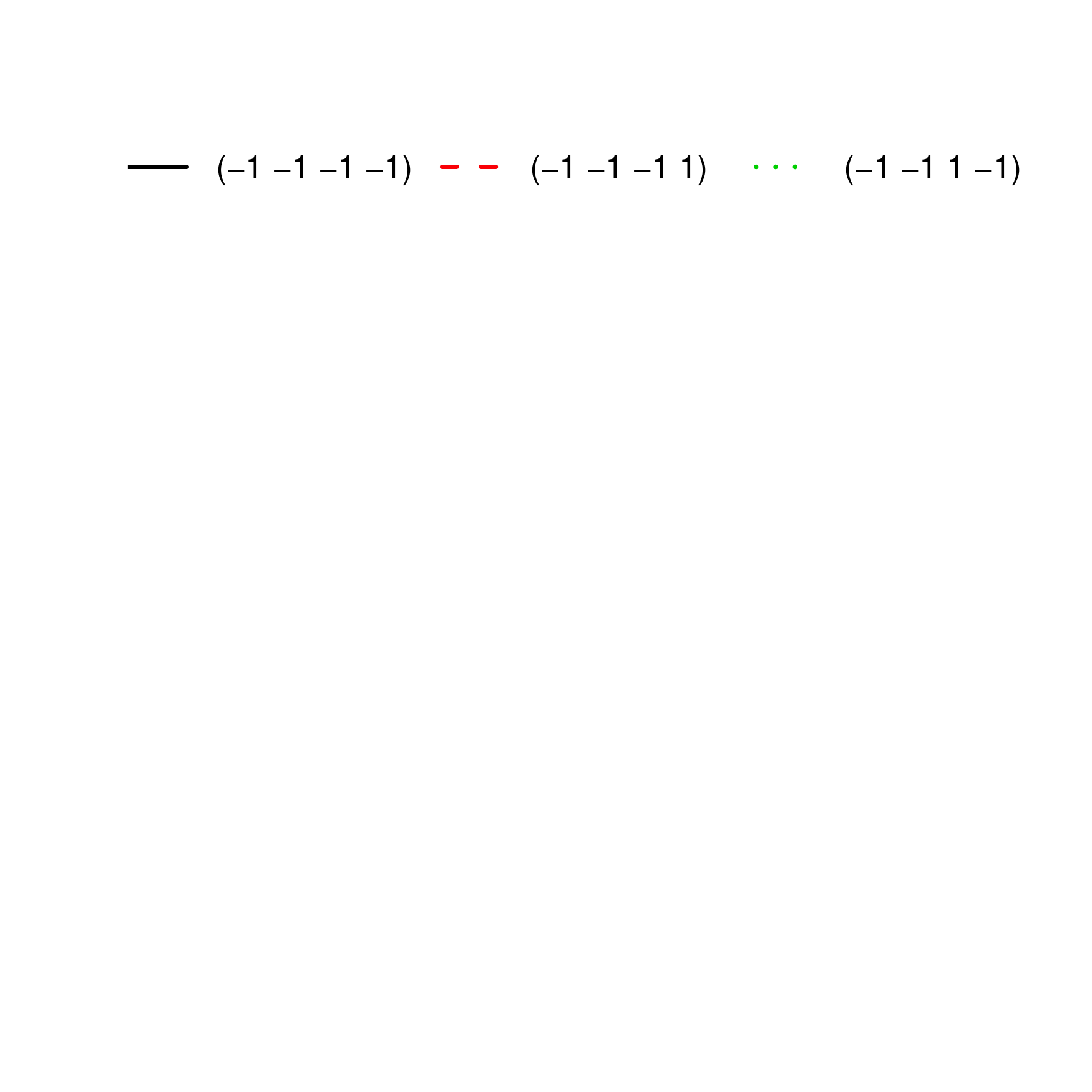}\\
\vspace{-2.85in}\includegraphics[width=3in]{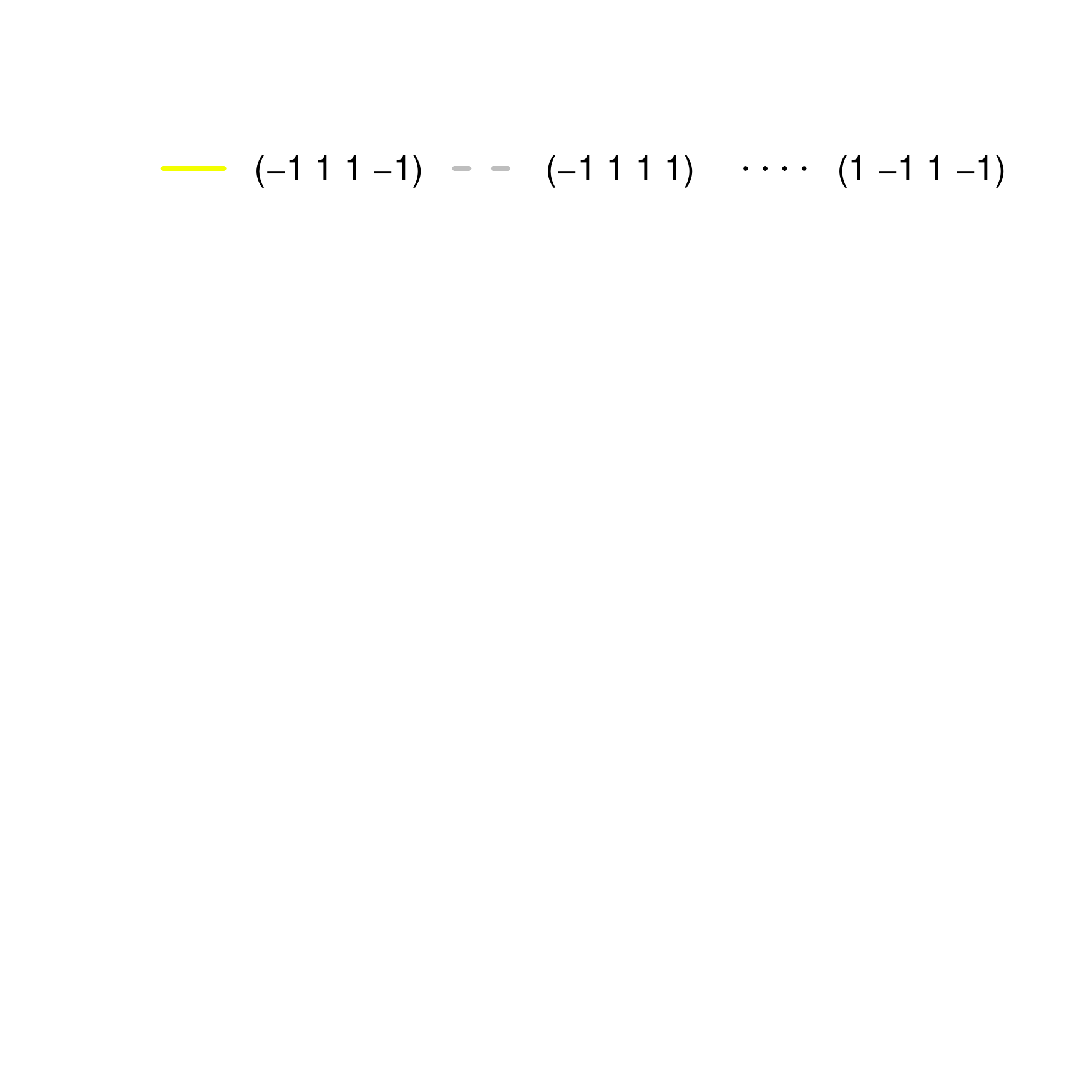}\\
\vspace{-2.4in}\includegraphics[width=3in]{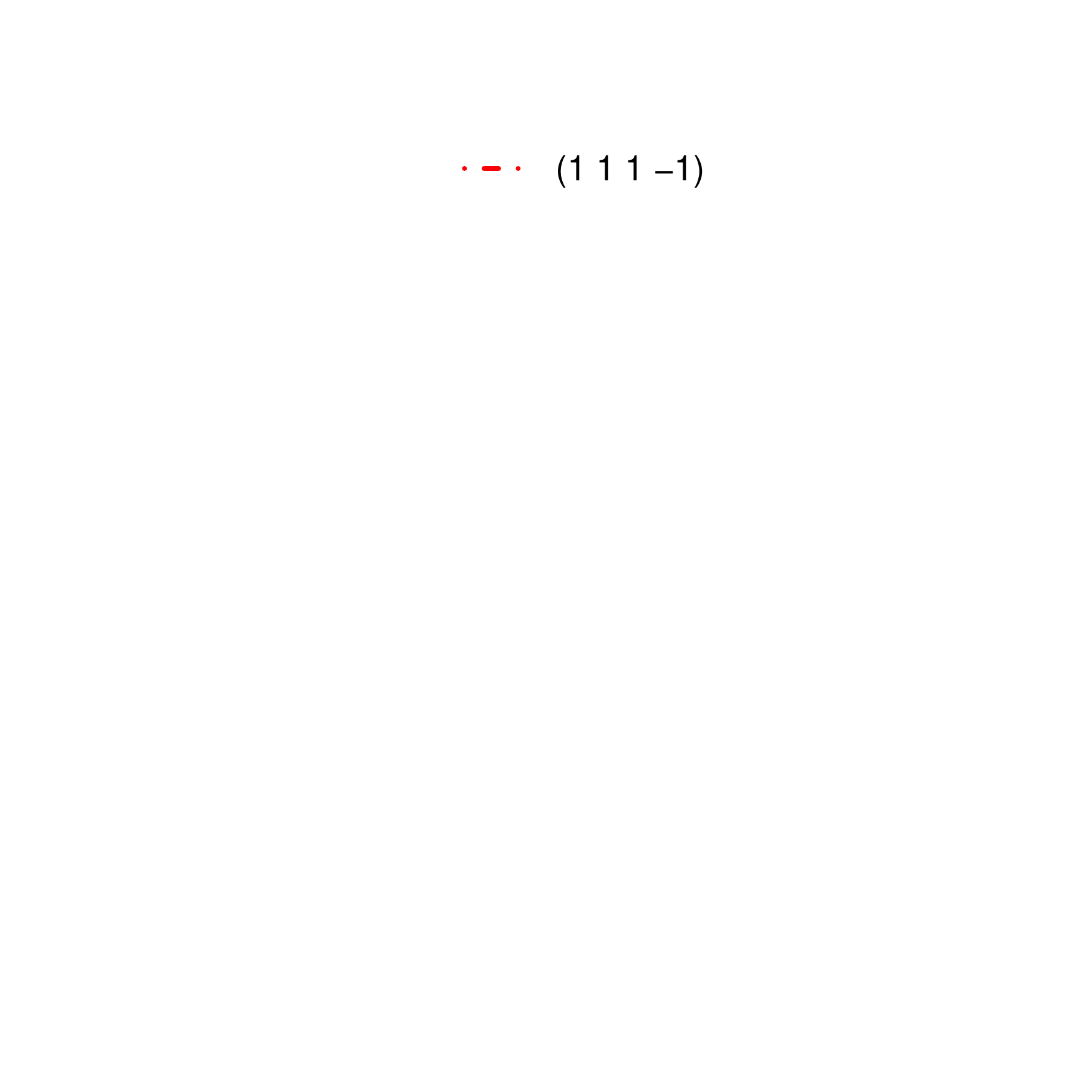}\\
\hline
\end{tabular}
\end{minipage}

&

\begin{minipage}{0.5\textwidth}
  \begin{tabular}{|ccccc|c|}
  \hline
A & B & ESD & Pulse & Voltage & $p_i$\\
$-1$ & $-1$ &$-1$ & $-1$ & $25.00$ & $0.0765$\\ 
$-1$ & $-1$& $-1$ & $-1$ & $27.55$ & $0.0133$\\ 
$-1$ & $-1$ & $-1$ &  $\phantom{-}1$ & $28.68$ & $0.0731$ \\ 
$-1$ & $-1$ & $-1$ &  $\phantom{-}1$ & $25.00$ & $0.0355$ \\ 
$-1$ & $-1$ &  $\phantom{-}1$ & $-1$ & $25.00$ & $0.1164$\\ 
$-1$ & $-1$ &  $\phantom{-}1$ &  $\phantom{-}1$ & $25.00$ & $0.0855$\\ 
$-1$ & $\phantom{-}1$ & $-1$ & $-1$ & $29.06$ & $0.0056$ \\ 
$-1$ & $\phantom{-}1$ & $-1$ & $-1$ & $25.00$ & $0.0882$\\ 
$-1$ & $\phantom{-}1$ & $-1$ & $\phantom{-}1$ & $25.00$ & $0.1012$ \\ 
$-1$ & $\phantom{-}1$ & $\phantom{-}1$ & $-1$ & $25.00$ & $0.0344$\\ 
$-1$ &  $\phantom{-}1$ &  $\phantom{-}1$ &$-1$ & $32.77$ & $0.1312$\\
$-1$ &  $\phantom{-}1$ &  $\phantom{-}1$ &  $\phantom{-}1$ & $25.00$ & $0.0922$\\
$\phantom{-}1$ & $-1$ & $\phantom{-}1$ & $-1$ & $25.00$ & $0.0138$\\ 
$\phantom{-}1$ & $\phantom{-}1$ &  $\phantom{-}1$ & $-1$ & $25.00$ & $0.1330$\\ 
\hline
  \end{tabular}
\end{minipage}

\end{tabular}
\caption{The sensitivity plot for the PSO-generated design for the voltage experiment. \textbf{Table 9.} The PSO-generated design for the voltage experiment. }
\label{voltage_equiv}
\end{figure}

After running the experiment, estimates of $\mathbf{\hat{\beta}} = (-7.17, 1.53, -0.11, -0.01, 0.13, 0.28, 0.38)$ were obtained. We take nominal values of $\mathbf{{\beta}} = (-7.5, 1.50, -0.2, -0.15, 0.25, 0.35, 0.4)$. Instead of treating voltage as a discrete factor with 5 levels we allow it to be continuous and range from 25 to 45. We used PSO with 50 particles, 150 maximum iterations, convergence tolerance of 0.0001, minimum lower efficiency bound of 99\%, and 400 maximum resets to identify a $D$-optimal design for this experiment. The locally optimal design obtained using PSO has only 14 support points. It is presented with its sensitivity plot in Table 9 and Figure~\ref{voltage_equiv}.

The PSO-generated design has five distinct voltages tested, but these voltages are not similar to the voltages tested in the original experiment. Instead the voltages are kept around 25 and 30. We can check the sensitivity plot in Figure~\ref{voltage_equiv} to see that the design is $D$-optimal. The relative efficiency of the original 80 point design to the design identified by PSO is $32.85\%$, indicating that the design obtained using particle swarm is approximately three times as efficient in terms of the $D$-optimality criterion.


\addtocounter{table}{1}

\section{Irregular Design Spaces}
 The bulk of the optimal designs reported in the literature are on prototype design spaces; for example, when factors are continuous the default seems to be on the unit cubiod or for mixture experiments, the design space is the regular simplex. In practice, some studies have irregularly shaped design spaces and this poses additional mathematical problems in finding an analytical description of optimal design.  Such design problems seem to have been not well studied in the literature even though they appear frequently in real problems. Draper and Guttman (1986) considered design problems when both bias and variance errors are present and the design space is from a class of flexible regions. The aim of this subsection is to show that PSO is flexible and can be directly modified to find an optimal design for such problems.

 As an example, consider the plastic molding experiment described by Anderson and Whitcomb (2004). In this experiment, temperate and pressure were varied as plastic was injected into a mold. If the temperature and pressure fell too low, there would not be enough plastic  injected into the model; if they were too high, the opposite problem occurred with too much plastic entering and the mold became ruined by flashing.  Consequently, physical constraints were imposed: temperature was allowed to vary from $450^{\circ}$ to $460^{\circ}$ F and pressure was allowed to vary from 1000 to 1300 subject to the constraint that $5600 \leq 10 \times Temperature + Pressure \leq 5800$. We note that even with these conditions, there were still other problems that could arise during the plastic injecting process such as the development of air pockets in the product. The response for the study was whether or not the plastic product had the proper structure.

We model the relationship between the binary response and the two independent variables temperature and pressure using logistic regression with the goal of finding an optimal design to estimate the effect of temperature and pressure on whether or not the product comes out correctly. We are interested in $logit(\mu) = \beta_0 + \beta_1 x_1 + \beta_2 x_2$ where $x_1 \in [450, 460]$, $x_2 \in [1100, 1300]$, and $5600 \leq 10x_1 + x_2 \leq 5800$. The probability of the plastic having the proper structure is $\mu$ and the nominal values for the intercept, the effect of temperature, and the effect of pressure are $\mathbf{\beta} = (0.05, 0.003, 0.007)$, respectively.

\begin{table}
 \caption{\label{irregular_design_example} Locally $D$-optimal designs on irregular design spaces for the flashing example. The left panel shows the optimal design on the constrained design space and the right panel shows the optimal design on the unconstrained design space. }
  \centering
  \makebox{
   \begin{tabular}{|cc|c|}
  \hline
  Temp & Pressure & $p_i$\\
  \hline
  450 & 1100.00 & 0.334 \\
  460 & 1200.00 & 0.335 \\
  460 & 1000.00 & 0.331 \\
\hline
  \end{tabular}
  }
  \makebox{
   \begin{tabular}{|cc|c|}
  \hline
  Temp & Pressure & $p_i$\\
  \hline
  450 & 1000.00 & 0.3220 \\
  450 & 1265.56 & 0.1899 \\
  460 & 1000.00 & 0.3214 \\
  460 & 1262.13 & 0.1667 \\
\hline
  \end{tabular}
  }
  \end{table}

\begin{figure}
\begin{center}
\includegraphics[width=2.4in,height=2.4in]{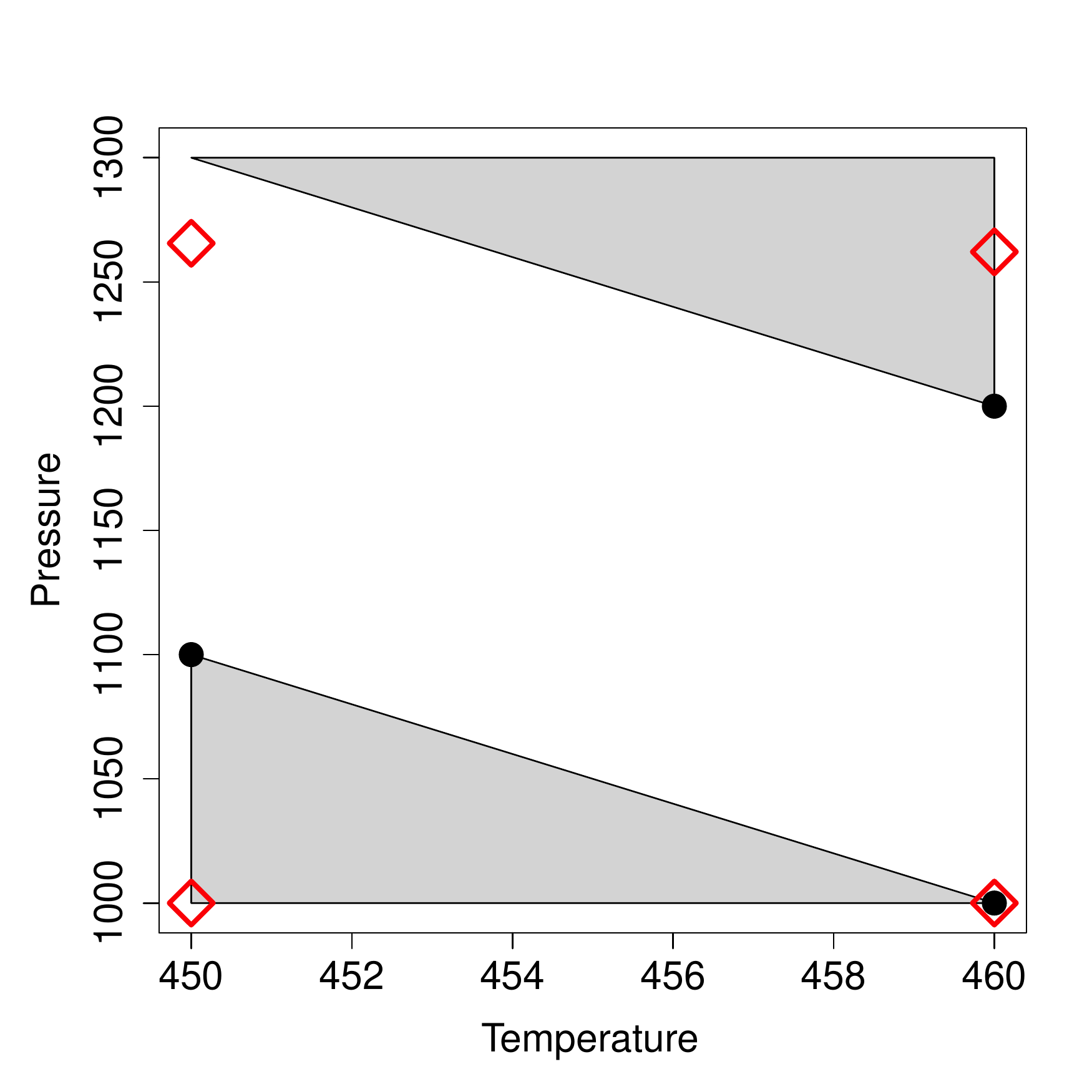}
 \end{center}
  \caption{\label{irregular_design_example_graphs} The black dots are the support points of the constrained locally $D$-optimal design on the irregular design space
(in white) and the diamonds are the support points of the locally $D$-optimal design on the larger rectangular design space.}
\end{figure}

The particle swarm algorithm can be easily modified  to accommodate additional constraints or complications in the optimization problem. In our case, all that is required is to modify the code to constrain  the swarm searches within the boundary of the user-specified space defined by the additional inequality constraints.  The left panel of Table~\ref{irregular_design_example} displays the resulting locally $D$-optimal design, which meets all the constraints on the irregularly design space.  Clearly this design is not the locally $D$-optimal design shown on the right panel, which does not meet the constraints specifications. Instead of running the algorithm until the equivalence theorem is satisfied we ran it 100 times with 25 particles and 200 maximum iterations until it converges (convergence tolerance 0.0001) and picked the best result.


\section{Summary and Conclusions}

In this paper we proposed a modified PSO algorithm to  find locally $D$-optimal designs for generalized linear models with a binary response under various link functions.  We demonstrated that PSO was able to find locally $D$-optimal designs for cases when there are all discrete factors, all continuous factors, and there is a mixture of discrete and continuous factors. Equivalence theorems were used to confirm the optimality of all PSO-generated designs among all designs on the given design space. We also demonstrated that PSO was able to find locally $D$-optimal designs for two real problems, as well as a problem with an irregular design space.

The PSO algorithm requires virtually no assumption on the optimization problem   and so it is applicable to situations where the model has multiple constraints on the design space or on its parameter space, or when the design criterion is non-differentiable. An appealing feature of PSO is that, unlike many other metaheuristic algorithms, it does not require users to carefully choose the tuning parameters for it to work effectively.  Many researchers across disciplines reported that the default tuning parameters for PSO work well, including those reported in Qiu et al. (2014) and Wong et al. (2015) who were among the first to apply PSO to solve optimal design problems.

PSO should and can find other types of optimal designs, including exact optimal designs for correlated responses or minimum bias designs. For such problems, there are no equivalence theorems to resort to because the design criteria are no longer convex or concave, and the only way PSO-generated designs can be confirmed is by theory, which is only feasible for relatively simple models. For example, \cite{Chen2015} were able to use PSO and generated exact $D$-optimal designs for the Michaelis-Menten model with correlated errors and confirmed their numerical results with the theoretical optimal designs available from \cite{Dette2014}.  In addition, \cite{Chen2015} were also able to use PSO and find optimal designs for more complicated settings than those considered in \cite{Dette2014}, where theoretical results are no longer available.  We close with the expectation that our work here can also be applied to healthcare studies with mixed factors, which are commonplace.  For example, Vago and Kemeny (2005) fitted a logistic model with 3 discrete variables (diabetes, sex, modality) and 3 continuous covariates (age, albumin and hemoglobin) using data from a clinical study conducted by Molnar, et al. (2005).  The binary outcome was whether patients experienced restless legs syndrome after renal transplantation.   Such studies can be expensive and a well-designed study can produce cost savings and accurate statistical inference at the same time.


\section{Acknowledgements}

The research of Dr. Mandal was in part supported by NSA Grant H98230-13-1-025. The research of Dr. Wong reported in this paper was partially supported by the National Institute of General Medical Sciences of the National Institutes of Health under Award Number R01GM107639. The contents in this paper are solely the responsibility of the authors and does not necessarily represent the official views of the National  Institutes of Health.

\bibliographystyle{rss}
\nocite{*}
\bibliography{optdbib}

\end{document}